\documentclass[showpacs,prb,floatfix,amsmath,amsfonts,superscriptaddress,twocolumn]{revtex4}
\usepackage{natbib,graphicx,amscd}
\usepackage[all,cmtip]{xy}
\usepackage[dvips]{changebar}
\usepackage[usenames]{color}
\usepackage{ulem}
\newcommand{\pd}[2]{\frac{\partial #1}{\partial #2}}
\newcommand{\pdd}[2]{\frac{\partial^2 #1}{\partial #2^2}}

\newcommand{\m}{\mathbf{m}}
\newcommand{\rhov}{\boldsymbol{\rho}}
\newcommand{\foe}{f_\textrm{oe}}
\newcommand{\thf}{\theta_{\textrm{f}}}
\newcommand{\z}{\mathbf{z}}
\newcommand{\Ic}{I_\textrm{c}}
\newcommand{\mf}{\mathbf{m}_\textrm{f}}
\newcommand{\M}{\mathbf{M}}
\newcommand{\Ms}{M_\textrm{s}}

\newcommand{\h}{\mathbf{h}}
\newcommand{\hoe}{\mathbf{h}_\textrm{oe}}
\newcommand{\he}{\mathbf{h}_\textrm{eff}}
\newcommand{\hk}{h_\textrm{k}}
\newcommand{\hkc}{h_\textrm{k}^\textrm{cr}}
\newcommand{\hkcm}{h_\textrm{k}^\textrm{cr,m}}
\newcommand{\sus}{\sigma_\textrm{sus}}
\newcommand{\susm}{\sigma_\textrm{sus}^\textrm{min}}
\newcommand{\wm}{\omega^\textrm{max}}
\newcommand{\ws}{\omega_\textrm{s}}
\newcommand{\wsp}{\omega_\textrm{s,1}}
\newcommand{\sis}{\sigma_\textrm{s}}
\newcommand{\ki}{k_\textrm{i}}
\newcommand{\ko}{k_\textrm{o}}
\newcommand{\pht}{\widetilde{\Phi}}
\newcommand{\eth}{\eta_\Theta}
\newcommand{\mzm}{m_z^\textrm{max}}
\newcommand{\ths}{\Theta_\textrm{s}}
\newcommand{\rdrop}{\rho_\textrm{drop}}
\newcommand{\phc}{\Phi_\textrm{c}}

\newcommand{\vf}{\boldsymbol{\varphi}}
\newcommand{\eps}{\varepsilon}

\newcommand{\R}{\mathbb{R}}
\newcommand{\x}{\mathbf{x}}
\newcommand{\rv}{\mathbf{r}}
\newcommand{\lex}{l_\textrm{ex}}
\newcommand{\dri}{d\rho_\textrm{in}}
\newcommand{\dro}{d\rho_\textrm{out}}

\newcommand{\rem}[1]{}

\DeclareMathOperator{\re}{Re}
\DeclareMathOperator{\im}{Im}

\begin{document}
\title{Theory for a dissipative droplet soliton excited by a spin torque nanocontact}
\date{\today}
\author{M. A. \surname{Hoefer}}
\affiliation{Department of Mathematics,
  North Carolina State University, Raleigh, NC 27695, USA}
\email{mahoefer@ncsu.edu}
\author{T. J. \surname{Silva}}
\author{Mark W. \surname{Keller}}
\affiliation{National Institute of Standards and Technology, Boulder,
  Colorado 80305, USA}
\thanks{Contribution of the U.S. Government, not subject to copyright.}
\begin{abstract}
  A novel type of solitary wave is predicted to form in spin torque
  oscillators when the free layer has a sufficiently large
  perpendicular anisotropy.  In this structure, which is a dissipative
  version of the conservative droplet soliton originally studied in
  1977 by Ivanov and Kosevich, spin torque counteracts the damping
  that would otherwise destroy the mode.  Asymptotic methods are used
  to derive conditions on perpendicular anisotropy strength and
  applied current under which a dissipative droplet can be nucleated
  and sustained.  Numerical methods are used to confirm the stability
  of the droplet against various perturbations that are likely in
  experiments, including tilting of the applied field, non-zero spin
  torque asymmetry, and non-trivial Oersted fields.  Under certain
  conditions, the droplet experiences a drift instability in which it
  propagates away from the nanocontact and is then destroyed by
  damping.
\end{abstract}

\maketitle

\section{Introduction}
\label{sec:introduction}

The concept of a soliton as a localized particle-like excitation that
preserves its shape can be extended to systems that are far from
thermodynamic equilibrium through the concept of a ``dissipative
soliton'' \cite{akhmediev_dissipative_2008}.  This allows us to
analyze a broad range of physical, chemical, and biological nonlinear
systems in which localized excitations are observed.  Driven magnetic
systems, especially those of technological interest, exhibit strongly
nonlinear dynamics and are an ideal experimental domain for exploring
the dissipative soliton model.

In this paper, we provide an analytical theory for a novel, localized
oscillation mode in a spin torque oscillator with a free layer having
perpendicular magnetic anisotropy. The salient features of this mode
include a frequency well below that of uniform ferromagnetic
resonance, a weak dependence of frequency on bias current, and a
precession angle at the maximal value of $90^\circ$.  Combining
numerical micromagnetic simulations with an asymptotic analysis of the
equations of motion, we identify this mode as a dynamic, dissipative
magnetic soliton that is closely related to the ``magnon droplet''
predicted by Ivanov and Kosevich in 1977
\cite{ivanov_bound-states_1977,kosevich_magnetic_1990}. The mode
central region exhibits  magnetization
pointing  nearly opposite to its equilibrium
direction  with a perimeter 
manifesting $90^\circ$ precession.  From our asymptotic
analysis, we derive conditions on perpendicular anisotropy and bias
current for the nucleation and existence of the dissipative droplet.
Using our numerical simulations, we analyze the stability of the
dissipative droplet soliton as a function of applied magnetic field,
bias current, and spin torque asymmetry.

Solitons in conservative systems occur when nonlinear terms in the
equation of motion balance the effects of dispersion
\cite{ablowitz_solitons_1981}.  A classic example is a light pulse
moving in a lossless optical fiber: the change of refractive index
with frequency (dispersion) tends to make the pulse spread out, but
for a certain pulse shape, the change of refractive index with light
intensity due to the optical Kerr effect (nonlinearity) exactly
balances the dispersion.  Pulses having this shape can propagate
without spreading and are called solitary waves or solitons.  The
balance between nonlinearity and dispersion typically allows for the
existence of a continuous family of solitons that can be excited in
the system, rather than a single solution.  In the optical fiber
example, the family can be parametrized, for example, by pulse
amplitude, and there is a continuous range of amplitudes that satisfy
the soliton balancing condition.

Dissipative solitons \cite{akhmediev_dissipative_2008} are
characterized by an additional balancing condition between gain and
loss that typically allows only a single solution for a given set of
external parameters. Although conservative soliton models can explain
weakly nonlinear behavior seen in magnetic systems of exceptionally
low damping \cite{zvezdin_contribution_1983}, damping is not a small
effect for many magnetic systems of both fundamental and technological
interest. By combining classical soliton theory with bifurcation
theory of nonlinear dynamics and concepts of self-organization
\cite{akhmediev__2008}, the dissipative soliton concept provides a
framework for describing a broad range of soliton-like behaviors.
Here, we apply this concept to a nanoscale ferromagnetic system in
which both damping and a driving force (spin torque) are important.

Spin torque
\cite{slonczewski_current-driven_1996,berger__1996,slonczewski_excitation_1999}
occurs when a current is driven through a structure with alternating
magnetic and nonmagnetic layers in which spin-dependent conductance at
the interfaces results in a spin-polarized electron flow. When the
polarized electrons enter a ferromagnetic layer whose magnetization
$\M$ is not collinear with the electron spins, the transmitted spins
are rotated toward $\M$ and the angular momentum absorbed by the
ferromagnetic layer is known as the spin torque. Typical devices have
two ferromagnetic layers through which current is driven: a thick
``fixed'' layer that determines the direction of electron
polarization, and a thin ``free'' layer whose orientation can be
readily changed by spin torque. For current of the appropriate
polarity, spin torque opposes the intrinsic damping torque in the
system, and currents above a threshold produce dynamic states in which
$\M$ of the free layer can be manipulated without applying a magnetic
field. This effect has been used to control switching of nanoscale
magnetic elements \cite{myers__1999}, with potential applications in
computer memory and data storage.  The effect has also been used to
produce coherent, frequency-tunable microwave oscillations
\cite{kiselev_microwave_2003} in a nanoscale device known as a spin
torque oscillator (STO), with potential applications in integrated
microwave circuits for mixing and active phase control. Recent reviews
cover the physics of spin torque \cite{ralph_spin_2008} and its
possible applications \cite{silva_developments_2008}.

The equations of motion for $\M$ in the presence of spin torque
(presented below) are inherently nonlinear, and their full solution
for a general case is often studied by use of numerical methods.
Analytical methods can sometimes be applied by invoking restrictions
such as high symmetry, spatially uniform $\M$ (the ``macrospin''
model), and small precession amplitude (small angle between $\M$ and
its equilibrium direction).  These restricted cases have been used to
explain experimental results with mixed success. The local nature of
spin torque allows it to drive large amplitude excitations in which
$\M$ varies on the scale of the magnetic exchange length (typically a
few nanometers), something that applied magnetic fields cannot do.  As
we show here, this regime of strongly nonlinear, strongly nonuniform,
sustained magnetodynamics is amenable to theoretical examination using
numerical and analytical approaches. This regime is also
experimentally accessible in STOs.

We note that a different type of magnetic soliton generated by spin
torque, called a spin wave ``bullet'', was predicted by Slavin and
Tiberkevich in 2005 to occur in the point-contact geometry with
magnetic films exhibiting in-plane oriented anisotropy and in-plane
applied magnetic field \cite{slavin_spin_2005}.  For this case, the
precession frequency decreases with increasing current, which can
result in localization if the frequency falls below the bottom of the
spin wave band at the ferromagnetic resonance (FMR) frequency.  The
weakly nonlinear bullet soliton is a solution to a Nonlinear
Schr\"{o}dinger type equation with third-order nonlinearity in the
excitation amplitude.  As such, its predicted experimental signature
consists of subtle shifts in microwave output frequency and threshold
current relative to that expected for a non-localized mode.  In
contrast, the droplet soliton studied here exhibits dramatic
differences in behavior from that of a non-localized mode. This is due
to the fact that it is a strongly nonlinear solution of the full
equations of motion, rather than simply a third order expansion.

Domain walls \cite{schryer_1974}, magnetic bubbles
\cite{nielsen_bubble_1976}, and vortices \cite{guslienko_2005} are
examples of well studied, strongly nonlinear, localized structures
that occur in magnetic materials.  The droplet differs from these
static structures in that it is inherently dynamic; the frequency of
spin precession within the droplet is always greater than zero.  In
this work, we focus on the two-dimensional (2D), non-topological
droplet, but we note that droplets in two and three-dimensions come in
topological flavors as well \cite{kosevich_magnetic_1990}.

We begin in the next section by presenting an asymptotic analysis of
the model equations for the dissipative droplet in a high-symmetry
case. We will also derive the droplet's frequency \emph{vs.} current
relation in this section.  Section \ref{sec:dropl-phys-pert} is
devoted to the study of droplets in physically realistic situations
incorporating the current-induced Oersted field as well as canting of
the applied field and fixed layer.
Section~\ref{sec:excitation-droplet} details experimentally accessible
nucleation conditions for a droplet that take advantage of a small
amplitude instability.  In Sec.~\ref{sec:discussion}, we discuss
possible extensions of the theory and we relate the droplet to other
excitations in thin magnetic films. We conclude in
Sec. \ref{sec:conclusion} with a summary of the droplet's unique
properties.  Appendices \ref{sec:modul-inst} and
\ref{sec:numerical-method} provide details of our stability
calculation and numerical method, respectively.

\section{Droplet in a Nanocontact}
\label{sec:diss-dropl-solit}

\begin{figure}
  \centering
  \includegraphics[width=0.95\columnwidth,clip=true,trim=0 100 165 0]{
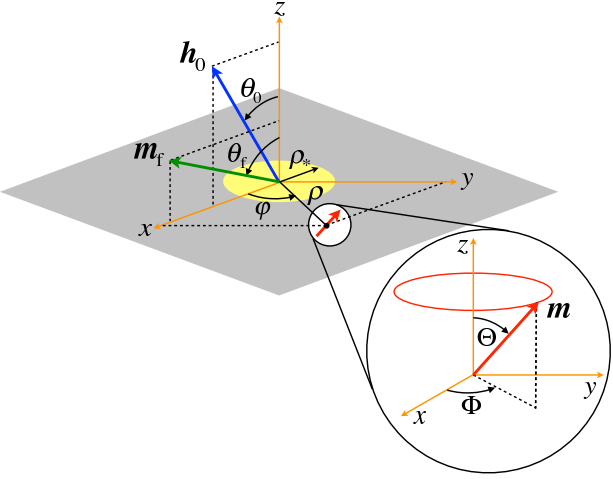}
  \caption{Schematic of nanocontact (yellow disk of radius $\rho_*$),
    the domain of $\m$ ($\R^2$ in polar coordinates with radius
    $\rho$, azimuthal angle $\varphi$), the range of $\m$ (unit sphere
    $\mathbb{S}^2$ with polar angle $\Theta$, azimuthal angle $\Phi$),
    orientations of the applied field $\mathbf{h}_0$ and fixed layer
    $\mf$.}
  \label{fig:coordinates}
\end{figure}

We consider the Landau-Lifshitz-Slonczewski equation in
non-dimensional
form\cite{slonczewski_current-driven_1996,ralph_spin_2008} describing
the free layer magnetization in polar coordinates ($\m =
\m(\rho,\varphi,t)$; bold symbols represent vectors in $\R^3$ or
$\R^2$, contextually dependent; e.g., $\m = [m_x,m_y,m_z]$)
\begin{equation}
  \label{eq:1}
  \begin{split}
    \pd{\m}{t} &= \underbrace{- \m \times \he}_\textrm{precession}
    \underbrace{- \alpha \m \times (\m \times \he)}_\textrm{damping}
    \\
    &\underbrace{+ \frac{\sigma V(\rho)}{1 + \nu \m \cdot \mf} \m
      \times (\m \times \mf)}_\textrm{spin torque}, \quad \m : \R^2
    \to
    \mathbb{S}^2, \\
    \he &= \nabla^2 \m + \h_0 + \hoe + (\hk - 1) m_z \z, \quad |\m|
    \equiv 1 .
  \end{split}
\end{equation}
Figure \ref{fig:coordinates} is a schematic of the nanocontact
geometry and coordinate systems considered in this work.  The
magnetization $\m = \M/\Ms$ and fields (e.g., $\h_0 =
\mathbf{H}_0/\Ms$) are normalized by the saturation magnetization
$\Ms$, time is normalized by $2\pi/|\gamma| \mu_0 \Ms$ ($\gamma$ is
the gyromagnetic ratio and $\mu_0$ is the permeability of free space),
and space is normalized by the exchange length $\lex =
\sqrt{D/|\gamma| \mu_0 \Ms \hbar}$ ($D$ is the exchange parameter and
$\hbar$ is the reduced Planck's constant).  We consider a free layer
whose thickness $\delta$ is much smaller than the magnetic excitation
wavelengths, so that a 2D model with local dipolar fields is justified
\cite{gioia_micromagnetics_1997}.  The precessional term is driven by
the effective field $\he$ incorporating exchange $\nabla^2 \m$, a
uniform applied field $\h_0 = h_0 [ \sin \theta_0, 0, \cos \theta_0]$
($h_0 \ge 0$), the current-induced Oersted field $\hoe = -\foe(\rho)
\vf$ (the definition of $\foe$ is given in
Ref.~\onlinecite{hoefer_model_2008} and summarized in
Sec.~\ref{sec:oersted-field}), the perpendicular field $\hk m_z \z$
due to crystal anisotropy ($\hk > 1$), and the demagnetizing field for
a thin film $-m_z \z$. The Landau-Lifshitz damping coefficient is
$\alpha$.  The spin torque term involves $\sigma = I / \Ic$, where $I$
is the current, $\Ic = (\lambda^2 + 1) 2 \Ms^2 e \mu_0 \pi r_*^2
\delta/(\hbar P \lambda^2)$ ($P$ and $\lambda \ge 1$ are the
polarization and spin torque asymmetry parameters, respectively
\cite{slonczewski_current-driven_1996}, $e$ is the electron charge,
$r_*$ is the nanocontact radius), $\nu = (\lambda^2 - 1)/(\lambda^2 +
1)$ ($0 \le \nu < 1$), $V(\rho) = \mathcal{H}(\rho_* - \rho)$, where
$\mathcal{H}$ is the Heaviside step function, $\rho_* = r_*/\lex$ is
the nondimensional contact radius, and $\mf = [\sin \thf,0,\cos \thf]$
is the uniform fixed layer magnetization.

Numerical computations of Eq.~\eqref{eq:1} will be presented in
Sec.~\ref{sec:dropl-phys-pert} by means of a method discussed in
Appendix \ref{sec:numerical-method}.  For the rest of this section, we
will focus on a high-symmetry case where
\begin{equation}
  \label{eq:53}
  \textrm{high symmetry case:} \quad \theta_0 = 0, \quad \hoe \equiv 0 .
\end{equation}
For the analysis of this case, it is helpful to consider
Eq.~\eqref{eq:1} in spherical coordinates $\m = [\cos \Phi \sin
\Theta, \sin \Phi \sin \Theta, \cos \Theta]$ (see
Fig.~\ref{fig:coordinates})
\begin{align}
  \label{eq:2}
  \pd{\Theta}{t} &= F[\Theta,\Phi] - \alpha G[\Theta,\Phi] 
   + \sigma V(\rho) P_\Theta[\Theta,\Phi], \\
  \label{eq:3}
  \sin \Theta \pd{\Phi}{t} &= G[\Theta, \Phi] + \alpha F[\Theta, \Phi]
   + \sigma V(\rho) P_\Phi[\Theta,\Phi] ,
\end{align}
where
\begin{align*}
  F[\Theta,\Phi] &= \sin \Theta \nabla^2 \Phi + 2 \cos \Theta \nabla
  \Phi \cdot \nabla \Theta, \\
  G[\Theta,\Phi] &= -\nabla^2 \Theta + \frac{1}{2} \sin 2\Theta
  (|\nabla \Phi |^2 + \hk - 1 ) \\
  \nonumber& + h_0 \sin \Theta, \\
  P_\Theta[\Theta,\Phi] &= \frac{-\cos \Theta \cos \Phi \sin \thf +
    \sin \Theta \cos \thf}{1 + \nu (\cos \Phi \sin \Theta \sin \thf +
    \cos \Theta \cos \thf)}, \\
  P_\Phi[\Theta,\Phi] &= \frac{\sin \Phi \sin \thf}{1 + \nu (\cos \Phi
    \sin \Theta \sin \thf + \cos \Theta \cos \thf)}.
\end{align*}
The polar angle $\Theta$ satisfies $0 \le \Theta \le \pi$, while the
azimuthal angle $\Phi$ is interpreted modulo $2 \pi$.

In the symmetric case of Eq.~\eqref{eq:53}, we can remove the explicit
dependence on $\hk$ from Eqs.~\eqref{eq:2} and \eqref{eq:3} by
introducing the following rescaling:
\begin{equation}
  \label{eq:52}
  \begin{split}
    \rho &= \rho'/\sqrt{\hk - 1}, \quad t = t'/(\hk - 1), \\
    \sigma &= (\hk - 1) \sigma', \quad h_0 = (\hk - 1) h_0', \quad
    \rho_* = \rho_*'/\sqrt{\hk - 1} .
  \end{split}
\end{equation}
Recall that we are assuming $\hk > 1$.  Then, with the scalings in
Eq.~\eqref{eq:52} and dividing Eqs.~\eqref{eq:2} and \eqref{eq:3} by
$\hk - 1$, we can, without loss of generality, take $\hk - 1 = 1$.
For the rest of this Section \ref{sec:diss-dropl-solit}, we will use
the scalings in Eq.~\eqref{eq:52} so that $\hk \to
2$.  

In this Section, we consider localized magnetization configurations
that satisfy
\begin{equation*}
  \lim_{\rho' \to \infty} \Theta(\rho',\varphi,t') = 0,
\end{equation*}
with sufficiently rapid decay.  As such, we define the magnetic energy
in terms of exchange and anisotropy energy via
\begin{equation*}
  \begin{split}
    \mathcal{E}[\Theta,\Phi] = \frac{1}{2} \int_{\R^2} \Big [
    &\underbrace{|\nabla \Theta|^2 + \sin^2 \Theta |\nabla \Phi
      |^2}_{\textrm{exchange}} \underbrace{+ \sin^2
      \Theta}_{\textrm{anisotropy}} \Big ] d \rv' .
  \end{split}
\end{equation*}
Note that the damping and spin torque terms break energy conservation
\begin{equation}
  \label{eq:22}
  \begin{split}
    \frac{d \mathcal{E}}{dt}&[\Theta,\Phi] = \\
    \int_{\R^2} \bigg \{
    &\sigma' \Big [(G[\Theta,\Phi] - h_0' \sin \Theta) P_\Theta[\Theta,\Phi]
    - F[\Theta,\Phi] 
    P_\Phi[\Theta,\Phi] \Big ] \\
    &- \alpha \Big [(G[\Theta,\Phi] - h_0' \sin \Theta)^2 
    + F[\Theta,\Phi]^2 \Big ] \bigg \} d\rv' .
  \end{split}
\end{equation}

\subsection{Conservative Droplet Soliton}
\label{sec:droplet-soliton}

In the absence of damping and spin torque ($\alpha = 0$, $\sigma' =
0$), Eqs.~\eqref{eq:2} and \eqref{eq:3} admit a continuous family of
exponentially localized, non-topological solutions known as magnon
droplet solitons
\cite{ivanov_bound-states_1977,kosevich_magnetic_1990}. These
solutions can be parametrized by the frequency $\omega_0$ and written
as
\begin{equation}
  \label{eq:55}
  \begin{split}
    \Theta &\equiv \Theta_0(\rho; \omega_0), \\
    \Phi &\equiv (\omega_0 + h_0') t', \\
    &= [\omega_0(\hk-1) + h_0] t .
  \end{split}
\end{equation}
They satisfy a balance between exchange (dispersion) and anisotropy
(nonlinearity) through the nonlinear eigenvalue problem
$F[\Theta_0,(\omega_0 + h_0') t'] = 0$, $G[\Theta_0,(\omega_0 + h_0')
t'] = (\omega_0 + h_0') \sin \Theta_0$ or
\begin{equation}
  \label{eq:6}
  \left ( \frac{d^2}{d \rho'^2} + \frac{1}{\rho'} \frac{d}{d \rho'}
  \right ) \Theta_0 -
  \frac{1}{2} \sin 2\Theta_0 + \omega_0 \sin \Theta_0 = 0,
\end{equation}
with the boundary conditions
\begin{equation}
  \label{eq:7}
  \frac{d \Theta_0}{d \rho'}(0;\omega_0) = 0, \quad \lim_{\rho' \to
    \infty} \Theta_0(\rho';\omega_0) = 0 .
\end{equation}
\begin{figure}
  \centering
  \includegraphics[width=0.8\columnwidth]{
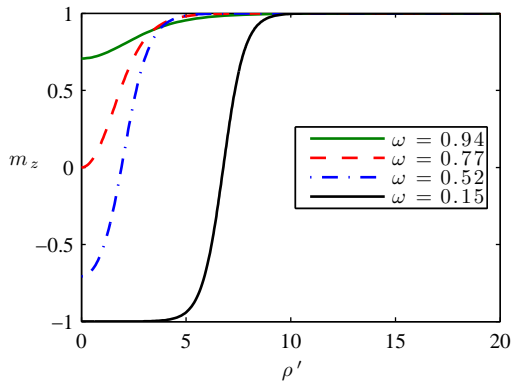}
  \caption{Conservative droplet profiles with $m_z = \cos \Theta_0$.}
  \label{fig:droplet_profiles}
\end{figure}
The conservative droplet solutions in Eqs.~(\ref{eq:55}),
\eqref{eq:6}, and \eqref{eq:7} have the following properties. The
polar angle $\Theta_0$ varies with radial distance and is independent
of time; thus the spatial distribution of $m_z = \cos\Theta_0$ is
static and azimuthally symmetric. The azimuthal angle $\Phi$ is
independent of position and varies linearly in time; thus all points
precess at a common frequency and in phase. It has been shown that
\cite{ivanov_bound-states_1977,kosevich_magnetic_1990} $\omega_0$
satisfies
 \begin{equation}
  \label{eq:mwk1}
   0 < \omega_0 < 1 .
\end{equation}
Therefore, the total precessional frequency $\omega_0 + h_0'$ varies
between the Zeeman frequency $h_0'$ and the frequency of spatially
uniform precession about $\Theta = 0$, $1 + h_0'$ (the FMR frequency).

While the conservative droplet does not have a closed-form analytical
expression, we calculate it by numerically solving Eq.~(\ref{eq:6})
subject to the boundary conditions (\ref{eq:7}) with the function
\texttt{bvp4c} in Matlab$^\textrm{\textregistered}$.  A plot of
several conservative droplet profiles is shown in
Fig.~\ref{fig:droplet_profiles}. We see that the amplitude at the
origin $1 - m_z(0;\omega_0) = 1 - \cos \Theta_0(0;\omega_0)$ decreases
as the frequency is increased.  While it may appear in
Fig.~\ref{fig:droplet_profiles} that $m_z(0;\omega_0 = 0.15) = 0$, in
fact all conservative droplets with $0 < \omega_0 < 1$ satisfy
$m_z(0;\omega_0) > -1$ owing to their non-topological structure.

The energy for the droplet $\mathcal{E}_0 \equiv
\mathcal{E}[\Theta_0,(\omega_0+h_0') t']$ satisfies
\cite{ivanov_bound-states_1977,kosevich_magnetic_1990}
\begin{equation}
  \label{eq:9}
  \frac{d \mathcal{E}_0}{d \omega_0} < 0.
\end{equation}
The fact that the energy is a decreasing function of frequency has
been used to argue that the conservative droplet is stable in a 2D
material \cite{kosevich_magnetic_1990}.  The 2D conservative droplet
embedded in an infinite, 3D magnet is known to be unstable.  However,
preliminary studies suggest that, for sufficiently thin films, the 2D
conservative droplet is stabilized.  This work is beyond the scope of
this paper and will appear elsewhere.

As we now show, the conservative droplet soliton can be generalized to
damped/driven systems such as a nanocontact. Whenever we refer to
``droplet'', we mean the dissipative droplet studied in the future
sections.  We will always use ``conservative droplet'' to describe the
solution of Eq.~(\ref{eq:6}) that is monotonically decaying to zero as
$\rho \to \infty$.

\subsection{Dissipative Droplet Soliton}
\label{sec:diss-dropl-solit-1}

We now extend the analysis of Kosevich, Ivanov, and Kovalev
\cite{ivanov_bound-states_1977,kosevich_magnetic_1990} to the case of
the dissipative droplet solution, where damping is no longer assumed
to be negligible, and spin torque is included in the analysis.  In
addition to the balance between exchange and anisotropy that was
required in the conservative droplet case, a further balance between
uniform damping (loss) and spatially localized spin torque (gain) will
be derived for the droplet to be sustained.  We will assume that the
spin torque and damping, while small, are not zero and are of the same
magnitude so that
\begin{equation*}
  \sigma = \mathcal{O}(\alpha), \quad 0 < \alpha \ll 1 .
\end{equation*}
We look for an asymptotic solution in the following form
\begin{equation}
  \label{eq:11}
  \begin{split}
    \Theta(\rv',t') &= \Theta_0(\rho';\omega) + \alpha \Theta_1(\rv',t') +
    \cdots, \\
    \Phi(\rv',t') &= (\omega + h_0') t' + \alpha
    \frac{\Phi_1(\rv',t')}{\sin 
      \Theta_0(\rho';\omega)} + \cdots .
  \end{split}
\end{equation}
We have set $\omega_0 \to \omega$ to distinguish the frequency of the
droplet $\omega$ from that of the conservative droplet $\omega_0$.  We
will conclude that there is no frequency shift due to damping and spin
torque so that $\omega = \omega_0$ here.  However, other perturbations
beyond those considered here could lead to a frequency shift.
Inserting the expansions \eqref{eq:11} into Eqs.~\ref{eq:2},
\ref{eq:3}, and equating like orders in $\alpha$ gives the following
equations for the perturbations $\Theta_1$ and $\Phi_1$:
\begin{align}
  \label{eq:12}
  \pd{\Theta_1}{t'} + L_\Phi \Phi_1 &= -(\omega + h_0') \sin \Theta_0
  \\
  \nonumber & \quad + \frac{\sigma'}{\alpha} V(\rho')
  P_\Theta[\Theta_0,(\omega + h_0') t'], \\ 
  \label{eq:13}
  \pd{\Phi_1}{t'} + L_\Theta \Theta_1 &= - \frac{\sigma'}{\alpha}
  V(\rho') P_\Phi[\Theta_0,(\omega + h_0')t'],
\end{align}
where the self-adjoint, Schr\"{o}dinger operators $L_\Phi$ and
$L_\Theta$ are
\begin{align*}
  L_\Phi &\equiv - \frac{\delta F}{\delta \Phi}[\Theta_0,(\omega + h_0')
  t'] \frac{1}{\sin \Theta_0} \\
  \nonumber &= -\nabla'^2 - \frac{d \Theta_0}{d \rho'}^2 + \cos^2
  \Theta_0 - \omega \cos \Theta_0, \\
  L_\Theta &\equiv \frac{\delta G}{\delta \Theta}[\Theta_0,(\omega +
  h_0') t']  - (\omega + h_0') \sin \Theta_0 \\
  \nonumber &= - \nabla'^2 + \cos 2 \Theta_0 - \omega \cos \Theta_0 .
\end{align*}
Note the following important property:
\begin{equation}
  \label{eq:8}
  L_\Phi \sin \Theta_0 = 0 .
\end{equation}

The rest of this section is concerned with the solution of the
perturbative equations (\ref{eq:12}) and (\ref{eq:13}) in two separate
cases.  We use standard solvability arguments for forced, linear
differential equations to remove secular terms (see
e.g.~Ref.~\onlinecite{kevorkian_multiple_1996}) in order to derive an
expression for the current $\sigma'$ at which the balancing condition
for the droplet is achieved.

\subsubsection{Case $\thf = 0$}
\label{sec:thf-=-0}

We first consider the case where the fixed layer is oriented perfectly
normal to the film plane so that $\thf = 0$.  In this regime, we can
study the effect of variable spin torque asymmetry $\nu$.

In addition to the scalings in Eq.~(\ref{eq:52}), we can also scale
out the dependence on the applied field $h_0'$ when $\thf = 0$ with
the substitution
\begin{equation}
  \label{eq:54}
  \Phi = \Phi' + h_0' t' .
\end{equation}
Then, $\Theta$ and $\Phi'$ satisfy Eqs.~(\ref{eq:2}) and (\ref{eq:3}),
as before, but with $h_0' = 0$.  This transformation shows that the
vertical applied field simply shifts the precessional frequency by
$h_0'$.  

We seek a solution to Eqs.~(\ref{eq:12}) and (\ref{eq:13}) in the form
$\thf = 0$, $h_0' = 0$, $\Theta_1 \equiv 0$, and $\Phi_1'(\rv',t') =
\Phi_1'(\rho')$ (i.e., linear phase evolution with time) which results
in the following non-homogeneous equation for $\Phi_1'$:
\begin{equation}
  \label{eq:17}
  \begin{split}
    L_\Phi \Phi'_1 &= -\omega \sin \Theta_0 + \frac{\sigma' V(\rho')
      \sin \Theta_0}{\alpha(1 + \nu \cos
      \Theta_0)}, \\
    \Phi_1'(0) &= 0, \quad \frac{d \Phi_1'}{d \rho'}(0) = 0 .
  \end{split}
\end{equation}
This equation is solvable if and only if the nonhomogeneous terms are
orthogonal to the kernel of $L_\Phi$.  By use of Eq.~(\ref{eq:8}),
multiplying the right-hand side of Eq.~(\ref{eq:17}) by $\rho \sin
\Theta_0$ and integrating from $0$ to infinity we obtain the existence
condition for a dissipative droplet, $\sigma' = \sus(\omega)$, where
\begin{equation}
  \label{eq:18}
  \sus(\omega) = \alpha \omega \frac{\int_0^\infty \sin^2
    \Theta_0(\rho';\omega) \rho' \, d
    \rho'}{\int\limits_0^\infty
    {\displaystyle \frac{ 
        V(\rho') \sin^2 \Theta_0(\rho';\omega) }{1 + \nu \cos
        \Theta_0(\rho';\omega)}} \rho' \, d \rho'} . 
\end{equation}
The choice $\sigma' = \sus(\omega)$ singles out a specific value for
the current as a function of the droplet frequency $0 < \omega < 1$.
This value of the current provides the necessary balance between
damping and spin torque, in addition to the balance between exchange
and anisotropy, in order to sustain the droplet.  Therefore, we call
$\sus(\omega)$ the \emph{sustaining current}.

\begin{figure}
  \centering
  \includegraphics[width=0.95\columnwidth]{
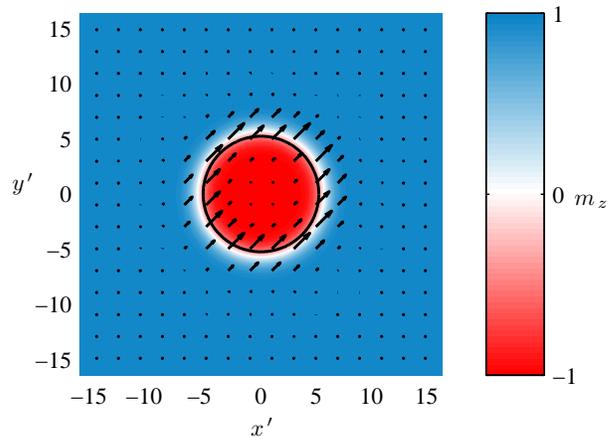}
  \caption{Dissipative droplet soliton for the high-symmetry case with
    $\theta_f = 0$.  The color scale corresponds to $m_z$, while the
    vector field corresponds to the in-plane components $(m_x,m_y)$.
    The circle here and in future plots represents the boundary of the
    nanocontact.  Parameters are $\sus/\alpha = 0.94$, $\omega =
    0.17$, $\rho_*' = 5.24$, and $\nu = 0$.}
  \label{fig:droplet_symmetric_contour}
\end{figure}

We can also understand the choice for the sustaining current
(\ref{eq:18}) by computing the rate of change in the energy for the
droplet from Eq.~(\ref{eq:22}):
\begin{equation}
  \label{eq:20}
  \begin{split}
    \frac{d \mathcal{E}}{d t'}[\Theta_0, \omega t' + &\alpha
    \Phi_1'/\sin \Theta_0]  = \\
    \omega \int_{\R^2} \Big [ &\sus(\omega) V(\rho') \frac{\sin^2
      \Theta_0}{1 + \nu \cos \Theta_0} \\
    & - \alpha \omega \sin^2 \Theta_0 \Big ] d\rv' +
    \mathcal{O}(\alpha^2) = \mathcal{O}(\alpha^2).
  \end{split}
\end{equation}
Therefore, the energy of the droplet is approximately conserved for
the choice $\sigma' = \sus$.  Figure
\ref{fig:droplet_symmetric_contour} shows a snapshot in time of a
dissipative droplet.

\begin{figure}
  \centering
  \includegraphics[width=0.8\columnwidth]{
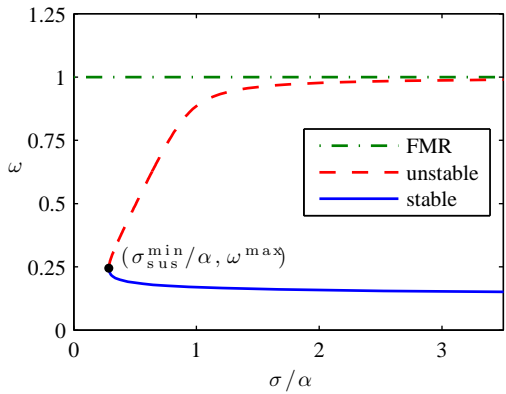}
  \caption{Dissipative droplet frequency (dashed and solid curves) as
    a function of sustaining current from Eq.~(\ref{eq:18}).  Parameters are
    $\rho_*' = 5.24$ and $\nu = 0$.  The dash-dotted line is the FMR
    frequency.}
  \label{fig:sustaining_current}
\end{figure}

Figure \ref{fig:sustaining_current} represents the numerical
evaluation of Eq.~(\ref{eq:18}) and shows the dependence of $\sus$ on
$\omega$.  The droplet frequency has two branches as the sustaining
current is varied, but only one branch is stable.  For a given droplet
frequency $\omega$, consider increasing the current slightly above the
sustaining value, $\sigma' = \sus(\omega) + \delta \sigma'$. From
Eq.~\eqref{eq:20}, the droplet energy will increase, and from
Eq.~\eqref{eq:9}, this increase in energy corresponds to a
\emph{decrease} in droplet frequency for a stable droplet
\cite{ivanov_bound-states_1977,kosevich_magnetic_1990}. Thus, the
upper branch in Fig.~\ref{fig:sustaining_current}, for which
increasing current causes an \emph{increase} in frequency, is
unstable.


Figure \ref{fig:sustaining_current} shows that there is a minimum
sustaining current $\susm$, maximum frequency $\wm$, and minimum $m_z$
in the center of the droplet $\mzm$ where
\begin{equation*}
  \begin{split}
    \susm &\equiv \min_{\omega \in (0,1)} \sus(\omega) \equiv \sus(\wm) ,\\
    \mzm &\equiv \cos \Theta_0(0;\wm) .
  \end{split}
\end{equation*}

\begin{figure}
  \centering
  \includegraphics[width=0.8\columnwidth]{
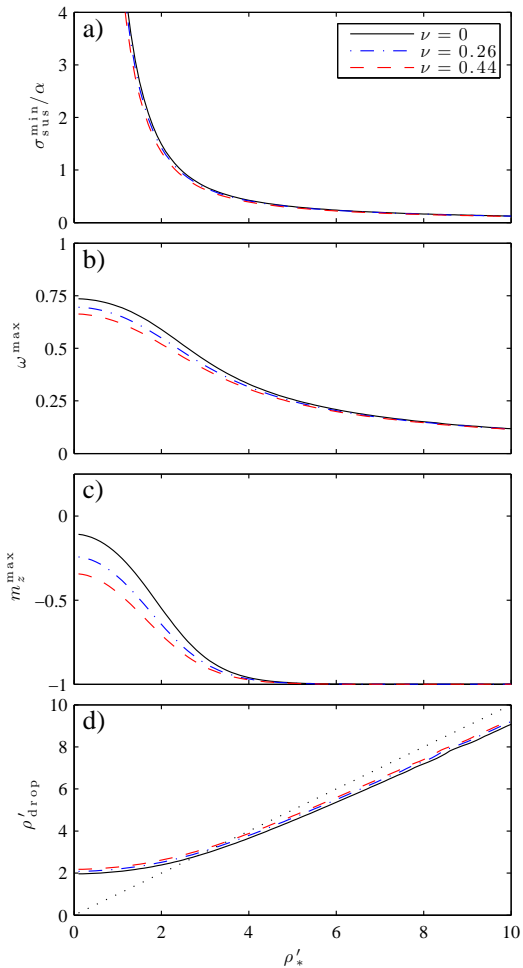}
  \caption{Dissipative droplet properties for varying contact radius
    $\rho_*'$ and spin torque asymmetry $\nu$: (a) minimum sustaining
    current, (b) maximum frequency, (c) maximum $m_z$ at origin, and
    (d) droplet radius (dotted line is $\rdrop' = \rho_*'$, plotted
    for comparison).}
  \label{fig:sus_min}
\end{figure}

While the specific choice of spin torque cutoff function $V(\rho')$
(e.g.~, we could have used a smooth cutoff, as opposed to a sharp,
Heaviside cutoff) may be important in numerical applications
\cite{berkov_magnetization_2007}, it has only a slight effect on the
droplet sustaining current.  For the Heaviside cutoff considered here,
Eq.~(\ref{eq:18}) simplifies to
\begin{equation}
  \label{eq:56}
  \sus(\omega) = \alpha \omega \frac{\int_0^\infty \sin^2
    \Theta_0(\rho';\omega) \rho' \, d
    \rho'}{\int\limits_0^{\rho_*'}
    {\displaystyle \frac{ 
        \sin^2 \Theta_0(\rho';\omega) }{1 + \nu \cos
        \Theta_0(\rho';\omega)}} \rho' \, d \rho'} .
\end{equation}
Equation (\ref{eq:56}) reveals the explicit dependence of the
sustaining current on two key spin torque parameters: the contact
radius $\rho_*' > 0$ and the spin torque asymmetry $0 \le \nu < 1$.
We now investigate properties of the dissipative droplet as a function
of these two parameters.  Figures \ref{fig:sus_min}(a-d) show the
dependence of $\susm$, $\mzm$, $\wm$, and the droplet radius $\rdrop'$
on the contact radius for various choices of the spin torque
asymmetry.  The droplet radius is defined to be the radius at half
maximum:
\begin{equation}
  \label{eq:19}
  m_z(\rdrop') = \frac{1}{2}(1 - \mzm) .
\end{equation}
Figure \ref{fig:sus_min}(a) shows that droplets excited by smaller
contacts require larger sustaining currents.  The dependence of
$\sus/\alpha$ on $\rho_*'$ and $\nu$ for small contact radii can be
made explicit by an asymptotic evaluation of the denominator in
eq.~(\ref{eq:56}) giving
\begin{equation}
  \label{eq:5}
  \frac{\sus(\omega)}{\alpha} = \frac{2 \omega [1 + \nu \cos
    \Theta_0(0;\omega)] \int_0^\infty \sin^2 \Theta_0(\rho;\omega)
    \rho d \rho}{\rho_*'^2 \sin^2 \Theta_0(0;\omega)} + \mathcal{O}(1), \quad 0 <
  \rho_*' \ll 1 .
\end{equation}
Therefore, $\susm/\alpha$ grows like $1/\rho_*'^2$ for small contact
radii in agreement with Fig.~\ref{fig:sus_min}(a).  Interestingly, we
find that the droplet is constrained to have a frequency $0 < \omega
\lesssim 0.75$ (Fig.~\ref{fig:sus_min}(b)), significantly below the
FMR frequency of 1.  Also, since $\mzm < 0$
(Fig.~\ref{fig:sus_min}(c)), the droplet always has some region near
its center that is partially inverted with respect to the surrounding
magnetization.  As shown in Fig.~\ref{fig:sus_min}(d), the droplet is
well localized within the nanocontact; e.g.~$\rdrop' < \rho_*'$, when
$\rho_*' \gtrsim 3$.  There is a minimum droplet radius of about $2
\sqrt{\hk - 1} \,\lex$ in dimensional units.  Finally, spin torque
asymmetry has only a small, perturbative effect and does not
substantially alter the droplet solution.


\subsubsection{Case $\nu = 0$}
\label{sec:eta-=-0}

We consider now the case without spin torque asymmetry (i.e.~$\nu =
0$) in which case we can study the effect of varying the angle of the
fixed layer $\thf$ on the droplet dynamics.  In this section, we use
the rescalings in Eq.~(\ref{eq:52}) but do not use the transformation
(\ref{eq:54}).

We solve Eqs.~(\ref{eq:12}) and (\ref{eq:13}) with $\nu = 0$,
$\Theta_1(\rv',t') \equiv \ths(\rv') \sin \omega t'$, and
$\Phi_1(\rv',t') \equiv \widetilde{\Phi}(\rv') + \phc(\rv') \cos
\omega t'$, leading to the following system of non-homogeneous
equations:
\begin{align}
  \label{eq:27}
  L_\Phi \phc + \omega \ths &= -\frac{\sigma'}{\alpha} \sin \thf
  V(\rho') \cos \Theta_0, \\
  \label{eq:28}
  L_\Theta \ths - \omega \phc &= - \frac{\sigma'}{\alpha} \sin \thf
  V(\rho'), \\
  \label{eq:29}
  L_\Phi \pht &= - (\omega + h_0') \sin \Theta_0 +
  \frac{\sigma'}{\alpha} \cos \thf V(\rho') \sin \Theta_0 .
\end{align}
Note that, in contrast to the case $\thf = 0$, the overall phase
$\Phi$ no longer evolves linearly in time.  Applying $L_\Theta$ to
Eq.~(\ref{eq:27}) and $L_\Phi$ to Eq.~(\ref{eq:28}) gives the
decoupled system:
\begin{align*}
  (L_\Theta L_\Phi + \omega^2) \phc = &~\frac{\sigma'}{\alpha}
  \sin \thf [ \omega V(\rho') \\
  \nonumber &-
  L_\Theta \{ V(\cdot) \cos \Theta_0(\cdot) \}(\rho')] ,\\
  (L_\Phi L_\Theta + \omega^2) \ths = &~\frac{\sigma'}{\alpha}
  \sin \thf [ L_\Phi \{ V(\cdot) \}(\rho') \\
  \nonumber &- \omega V(\rho')
  \cos \Theta_0(\rho')] .
\end{align*}
These equations are always solvable if $L_\Theta L_\Phi + \omega^2$
and $L_\Phi L_\Theta + \omega^2$ are strictly positive operators.  One
can show that $L_\Phi \ge 0$, so that $L_\Phi L_\Theta + \omega^2 >
0$.  One can also show that $L_\Theta \ge - \eth^2$, where $-\eth^2$
is the smallest eigenvalue of $L_\Theta$ and is strictly negative.
Then, $L_\Theta L_\Phi \ge -\eth^2$.  We have verified by numerical
computation of $\eth$ that $\omega^2 > \eth^2$ so that $L_\Theta
L_\Phi + \omega^2 > 0$ as required, and Eqs.~(\ref{eq:27}) and
(\ref{eq:28}) are solvable.

We are interested in the solvability condition for $\pht$ in
Eq.~(\ref{eq:29}), which is
\begin{equation}
  \label{eq:30}
  \sus(\omega) = \alpha (\omega + h_0') \sec \thf \frac{\int_0^\infty \sin
    \Theta_0(\rho';\omega) \rho'\, d
    \rho'}{\int_0^\infty
    V(\rho') \sin^2 \Theta_0(\rho';\omega)  \rho' \, d \rho'} .
\end{equation}
Note that the applied field appears only as a shift in the droplet
frequency, as in the case $\thf = 0$ studied in the previous section.
The expression (\ref{eq:30}) for the sustaining current agrees with
the previously derived sustaining current in Eq.~(\ref{eq:18}) when
$\nu = 0$ and $\thf = 0$, as required.  Thus, the dominant effect of
rotating the fixed layer is to increase the sustaining current in
proportion to $\sec \thf$.

\section{Physical Perturbations of a Dissipative Droplet}
\label{sec:dropl-phys-pert}

So far, we have considered the dissipative droplet solution only for a
simplified geometry where asymptotic methods can be applied.  In these
cases, the external field is both uniform and oriented perfectly
perpendicular to the film plane. By making this geometrical
simplification, we were able to factor out the contribution of the
external field from the droplet solution.  However, in a real
point-contact system, the external field is neither uniform nor
perfectly perpendicular.  In particular, the current flowing through
the contact is an additional source of spatially inhomogeneous
magnetic field,  the
  Oersted field, and the applied uniform magnetic field in actual
experiments is usually tilted away from the perpendicular axis.  We
employ micromagnetic simulations to investigate how these physically
important perturbations to the external field alter the ideal droplet
solution.  The numerical details used for our simulations are
presented in Appendix~\ref{sec:numerical-method}.

We find that the combination of external field tilt and the Oersted
field breaks the symmetry of the solution such that the droplet is no
longer centered in the middle of the contact. As a result, the
solution takes on a nontrivial inhomogeneous phase structure where the
phase of the spin precession closer to the center of the contact
precedes the phase further from the center of the contact. In
addition, the spatial structure of the droplet is no longer perfectly
circular.  For some particular combinations of simulation parameters,
the droplet breaks away from the contact altogether and dissipates, a
behavior we call a drift instability. When this occurs, the droplet
may maintain its form for many precession cycles, but it eventually
decays, since outside the contact there is no spin torque excitation to
balance damping.

\subsection{Oersted Field}
\label{sec:oersted-field}

\begin{figure}
  \centering
  \includegraphics[width=0.8\columnwidth]{
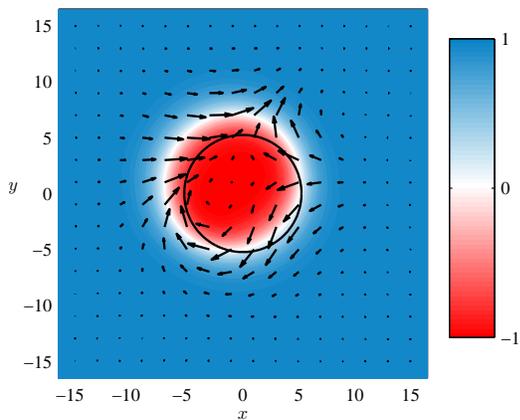}
  \caption{Dissipative droplet in the presence of the Oersted field
    and with the applied field and fixed layer nearly perpendicular to
    the plane..  Parameter values are $\hk = 1.25$, $\alpha = 0.03$,
    $\rho_* = 5.24$, $h_0 = 1.8$, $\theta_0 = 0.00001^\circ$, $\thf =
    0.40^\circ$, $\nu = 0.257$, $\sigma = 0.196$.}
  \label{fig:droplet_oe_perturb}
\end{figure}

First, we consider the effect of the current induced Oersted field
while keeping the applied field and fixed layer oriented almost normal
to the film plane (canted by $0.00001^\circ$ and $0.40^\circ$, resp.).
The reason for this slightly asymmetric configuration is to test
whether high-symmetry solutions are structurally stable to small
changes in the system parameters.  Such a configuration is
experimentally possible, in principle.  Our model for the Oersted
field was presented in Ref.~\onlinecite{hoefer_model_2008} and takes
the form
\begin{equation*}
  \hoe = - f_\textrm{oe}(\rho) \vf ,
\end{equation*}
where
\begin{equation}
  \label{eq:58}
  \begin{split}
    f_\textrm{oe}(\rho) &= g_\textrm{oe}(\rho) + \frac{I}{2 \pi \Ms r_*}
    \left \{
      \begin{array}{cc}
        \rho/\rho_*, & 0 < \rho < \rho_*, \\
        \rho_*/\rho, & \rho_* < \rho,
      \end{array} \right . .
  \end{split}
\end{equation}
The function $g(\rho)$ given in Ref.~\onlinecite{hoefer_model_2008}
involves integrals of Bessel functions and depends on the geometry of
the current-density distribution.  The parameters defining $g$ in
Ref.~\onlinecite{hoefer_model_2008} are $d = 1.67$, $z_* = -0.925$,
and $a = 2.92$.  The other, more dominant, term in Eq.~(\ref{eq:58})
is the field generated by an infinitely long conducting wire.  As an
example, for the simulation in Fig.~\ref{fig:droplet_oe_perturb}, we
have $\max_{\rho \in [0,\infty)} |g_\textrm{oe}(\rho)| = 0.0081$,
while $I/2\pi \Ms r_* = 0.086$, an order of magnitude difference.

Figure~\ref{fig:droplet_oe_perturb} illustrates how the Oersted field
changes the structure of the droplet. In contrast with the symmetric
case in Fig.~\ref{fig:droplet_symmetric_contour}, the azimuthal angle
$\Phi$ shows significant spatial variations.  The droplet is also
slightly shifted off-center.  The strong phase variations are
indicative of a tendency for the droplet to propagate
\cite{piette_localized_1998}.  In some cases, although not for the
simulation in Fig.~\ref{fig:droplet_oe_perturb}, the droplet breaks
free from the nanocontact.  This drift instability is discussed
further in Sec.~\ref{sec:canted-applied-field}.

\begin{figure}
  \centering
  \includegraphics[width=0.8\columnwidth]{
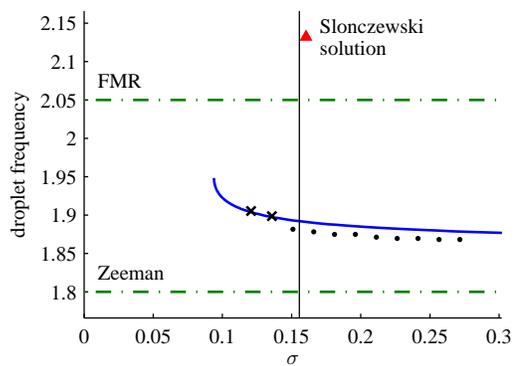}
  \caption{Dots: perturbed  droplet frequency as a function
    of current with the Oersted field computed from micromagnetic
    simulations.  Crosses: perturbed 
     droplets that undergo a drift instability; the
    frequency is calculated before the instability manifests.  Solid
    curve:  droplet frequency as a function of the
    sustaining current from Eq.~(\ref{eq:56}).  Dash-dotted: the
    Zeeman ($h_0$) and FMR ($h_0 + \hk -1$) frequencies.  Triangle:
    the Slonczewski critical current and onset frequency for high
    symmetry \cite{slonczewski_excitation_1999} (see Appendix
    \ref{sec:modul-inst}).  Solid vertical line: numerically computed
    threshold current in the presence of the Oersted field and the canted fixed
    layer.  Parameter values are the same as in
    Fig.~\ref{fig:droplet_oe_perturb}. 
  }
  \label{fig:droplet_f_I_oe}
\end{figure}

In Fig.~\ref{fig:droplet_f_I_oe} we compare the numerically computed
perturbed droplet frequency as a function of current with the
sustaining current from Eq.~(\ref{eq:56}).  We find that the droplet
frequency is approximately shifted down by an overall amount of
$0.012$, but otherwise follows the same trend as the symmetric result.
This behavior demonstrates that the analysis of the previous section,
despite the necessary high symmetry restrictions, yields relevant,
qualitative information about localized structures excited in a
physically realizable nanocontact.

As shown in Sec.~\ref{sec:excitation-droplet}, when the current
exceeds a threshold value; e.g., the Slonczewski critical
current\cite{slonczewski_excitation_1999}, a droplet can nucleate for
sufficiently large anisotropy.  The vertical line in
Fig.~\ref{fig:droplet_f_I_oe} is the numerically computed threshold
current.  It differs from Slonczewski's result because $\hoe \ne 0$,
$\theta_0 > 0$, and $\thf > 0$.  This threshold for droplet nucleation
suggests a hysteretic effect that will be discussed in
Sec.~\ref{sec:hysteresis}.

Our micromagnetic simulations also show that the perturbed droplet,
for certain parameter choices (e.g., the crosses in
Fig.~\ref{fig:droplet_f_I_oe}), undergoes a drift instability.  This
behavior will now be investigated further.


\subsection{Canted Applied Field, Fixed Layer, and Oersted Field}
\label{sec:canted-applied-field}

\begin{figure*}
  \centering
  \includegraphics[width=2\columnwidth]{
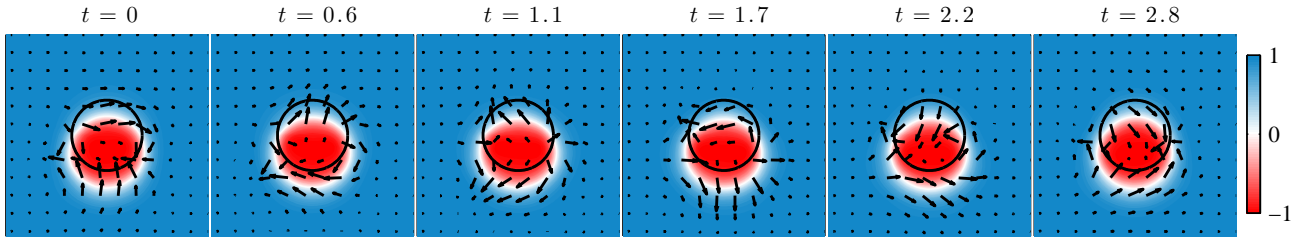}
   \caption{Time sequence of a strongly perturbed droplet over one
     period of precession in the presence of a canted applied field,
     canted fixed layer, and Oersted field.  Parameter values are $\hk
     = 1.25$, $\alpha = 0.03$, $\rho_* = 5.24$, $h_0 = 1.8$, $\theta_0
     = 5^\circ$, $\thf = 31.4^\circ$, $\nu = 0.257$, $\sigma = 0.189$.
     The time of the initial panel here and in
     Fig.~\ref{fig:droplet_drift} is set to 0 for comparison.  The
     simulation actually began earlier.}
  \label{fig:droplet_strong_perturb}
\end{figure*}

In this section, we investigate the combined effects of the Oersted
field as well as canting of the applied field and fixed layer.  Figure
\ref{fig:droplet_strong_perturb} is a time sequence showing the
evolution of a strongly perturbed droplet over one precessional
period.  In contrast to the nearly symmetric configuration of
Fig.~\ref{fig:droplet_oe_perturb}, where the droplet was slightly
shifted to the left, the droplet is slightly shifted down, toward the
region of lower in-plane field.

\begin{figure*}
  \centering
  \includegraphics[width=2\columnwidth]{
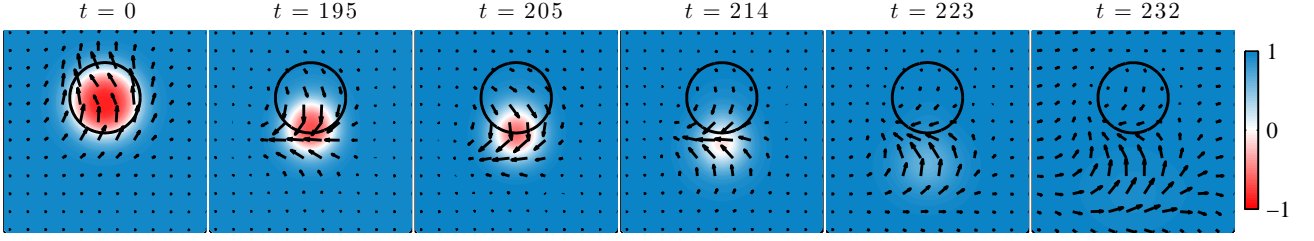}
  \caption{Time sequence showing the droplet drift instability for the
    same parameter values as in Fig.~\ref{fig:droplet_strong_perturb}
    but with smaller current $\sigma = 0.121$.  To facilitate
    visualization, the length of the in-plane magnetization vectors is
    normalized to the largest value in each frame.}
  \label{fig:droplet_drift}
\end{figure*}

As the system parameters are changed, the shifting of the droplet
center can be large enough to actually dislodge the droplet from the
nanocontact.  An example of this drift instability is shown in the
panels of Fig.~\ref{fig:droplet_drift}.  The current was taken to be
less than the current for Fig.~\ref{fig:droplet_strong_perturb}, which
did not experience a drift instability.  A number of precessional
periods pass before the droplet breaks free.  Once free, it
propagates, but because it no longer satisfies the required balancing
condition between damping and spin torque, it loses amplitude and
decays.  Once the droplet has drifted outside of the nanocontact, a
new one is formed if the nucleation conditions are satisfied (see
Sec.~\ref{sec:excitation-droplet}).  We have also observed droplets
that drift several nanocontact diameters before decaying, i.e.~the
central magnetization lifts up so that $\min_{\rv \in \R^2} m_z(\rv,t)
> 0$.  The manifold in parameter space in which the droplet manifests
a drift instability appears to be complicated.  Nevertheless, we
readily find parameter regimes where the droplet apparently does not
experience the drift instability, as in
Figs.~\ref{fig:droplet_oe_perturb}, \ref{fig:droplet_f_I_oe}, and
\ref{fig:droplet_strong_perturb}.

Notice that the droplet propagates down, in the $-\mathbf{y}$
direction.  Recall that the canting direction of the applied field is
along $\mathbf{x}$.  Due to the symmetry of the Oersted field, the
direction of propagation of the drifting droplet appears to track the
azimuthal angle of the applied field minus $90^\circ$, if $\theta_0$
is sufficiently large.  For example, if the applied field is canted
along $\mathbf{y}$, then the droplet will drift along $\mathbf{x}$ if
unstable.  This can be understood as a consequence of magnetostatic
interactions between the effective dipole moment of the droplet and
the field gradient associated with the Oersted field.  Given the
canting of the applied field, the effective dipole moment of the
droplet acquires an in-plane component that is drawn to the edge of
the contact where the Oersted field is also in the $-\x$ direction
such that the Oersted field gradient acts to trap the droplet.  In the
case of Fig.~\ref{fig:droplet_oe_perturb}, where the applied field is
canted only $0.00001^\circ$, the droplet is observed to drift to the
left rather than down.  Therefore, the strength of the Oersted field
and the in-plane component of the applied field have a strong effect
on the existence and dynamics of a drift instability.

\section{Nucleation of a Dissipative Droplet}
\label{sec:excitation-droplet}

\begin{figure*}
  \centering
  \includegraphics[width=2\columnwidth]{
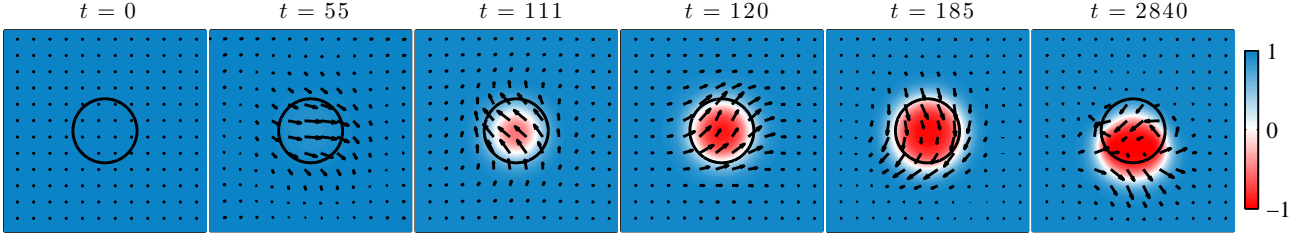}
  \caption{Birth of a dissipative droplet soliton for the current
    $\sigma = 0.186$, above the Slonczewski critical current $\sis =
    0.160$.  Parameter values are the same as those in
    Fig.~\ref{fig:droplet_strong_perturb}.}
  \label{fig:droplet_birth}
\end{figure*}

Figure \ref{fig:droplet_birth} shows the birth of the droplet pictured
in Fig.~\ref{fig:droplet_strong_perturb} starting from a state
pointing uniformly in the $\mathbf{z}$ direction.  For sufficiently
large perpendicular anisotropy $\hk$ and current, the magnetization
inside the nanocontact reverses and nucleates a droplet.  In this
section we show that the reversal mechanism is caused by an
instability of small amplitude waves.  We will study Eq.~(\ref{eq:1})
in the weakly nonlinear regime and find that the small amplitude
Slonczewski mode that exists near the threshold for the onset of
dynamics is stable/unstable depending on whether $\hk$ is less
than/greater than a critical value $\hkc$.  We find that $\hkc > 1$
due to exchange effects and converges to 1 as the contact size is
increased.  The value of $\hkc$ is important for the possible
experimental observation of a dissipative droplet.

\subsection{Stability Analysis of a Macrospin}
\label{sec:finite-dimens-model}

Before studying the PDE~(\ref{eq:1}), we consider the macrospin model
where spatial variation is neglected.  The stability analysis for this
model is suggestive and mathematically simpler.  However, we find that
the critical anisotropy field in the macrospin case $\hkcm$ satisfies
$\hkcm \le 1$ and depends strongly on the applied field, which
differs from the result obtained by analyzing the full PDE model,
where we find $\hkc > 1$ with weak applied field dependence.

We consider Eqs.~(\ref{eq:2}) and (\ref{eq:3}) neglecting all spatial
variation and inter-layer dipole coupling in the symmetric regime
$\theta_0 = 0$ and $\thf = 0$; i.e., we are assuming that the initial
condition for the system is in the parallel state.  In this case, the
equation for $\Theta$ is decoupled from $\Phi$, so we can just study
the scalar, first-order ODE
\begin{align}
  \label{eq:46}
  \dot{\Theta} &= - \alpha \sin \Theta [\cos \Theta (\hk - 1) + h_0] +
  \frac{\sigma \sin \Theta}{1 + \nu \cos \Theta} .
\end{align}
By linearizing Eq.~(\ref{eq:46}) about the equilibrium $\Theta \equiv
0$, we find that it becomes unstable when
\begin{equation*}
  \sigma > \sigma_0 \equiv \alpha (1 + \nu) (h_0 + \hk - 1) ,
\end{equation*}
in agreement with previous numerical and mathematical analyses of this
system contained in Ref.~\onlinecite{bazaliy_2004}.  We seek a
periodic equilibrium solution $\Theta(t) = \Theta_\textrm{e}$ just
above threshold by taking
\begin{equation*}
  \sigma = \sigma_0 + \eps, \quad 0 < \eps \ll \sigma_0 .
\end{equation*}
Then, we have
\begin{equation*}
  \Theta_\textrm{e} \sim \left [ \frac{2 \alpha \eps}{1 - \hk + \nu [2(1-\hk) -
      h_0]} \right ]^{1/2} .
\end{equation*}
This solution exists (is real valued) as long as
\begin{equation}
  \label{eq:4}
  \hk < \hkcm \equiv 1 - \frac{\nu h_0}{1 + 2\nu} .
\end{equation}
Furthermore, one can show that this equilibrium is stable.  Therefore,
when $\hk < \hkcm$, the equilibrium $\Theta \equiv 0$ undergoes a
supercritical Hopf bifurcation as the current $\sigma$ is increased
beyond $\sigma_0$.  When $\hk > \hkcm$, there is no periodic solution,
and the system switches from $\Theta \equiv 0$ to the fully reversed
state $\Theta_\textrm{e} = \pi$ when $\sigma$ exceeds $\sigma_0$.

\subsection{Stability Analysis of the Micromagnetic System}
\label{sec:stab-analys-infin}

We consider Eq.~(\ref{eq:1}) in the symmetric regime with $\theta_0 =
0$, $\hoe \equiv 0$, $\thf = 0$, and the substitutions
\begin{equation*}
  u = m_x + i m_y, \quad m_z = \sqrt{1 - |u|^2} \sim 1 - \frac{1}{2}
  |u|^2 ,
\end{equation*}
where $|u| \ll 1$.  Then $u$ approximately satisfies a complex
Nonlinear Schr\"{o}dinger type equation
\begin{equation}
  \label{eq:32}
  \begin{split}
    i \pd{u}{t} = &~(1 + i \alpha) \nabla^2 u - (h_0 + \hk - 1) u - i
    \alpha (h_0 + \hk - 1)u \\
    &+ i \frac{\sigma V(\rho)}{1 + \nu} u + \frac{1}{2} (\hk - 1) | u
    |^2 u \\
    &+ \frac{1}{2} ( u \nabla^2 |u|^2 - |u|^2 \nabla^2 u ) + i \alpha
    | \nabla u |^2 u \\
    &+ \frac{i}{2} \left [ \alpha(h_0 + 2\hk - 2) - \frac{\sigma
      V(\rho)}{(1+\nu)^2} \right ] |u|^2 u .
  \end{split}
\end{equation}
Similar to the macrospin case, when $\sigma$ is increased past a
threshold value, the zero solution becomes unstable.  The threshold,
critical current and onset frequency were found by Slonczewski as a
solution to a linear eigenvalue problem
\cite{slonczewski_excitation_1999} (see Eq.~(\ref{eq:34})).
Incorporating weak nonlinear effects,
Ref.~\onlinecite{hoefer_theory_2005} showed that a small amplitude
periodic solution exists as a modulation of the Slonczewski mode for
$\hk = 0$ and $\nu = 0$.  This time-periodic, weakly nonlinear mode
was found to be numerically stable.  In this section and Appendix
\ref{sec:modul-inst}, we extend these results to $0 \le \hk < \hkc$
and $0 \le \nu < 1$, where $\hkc$ is defined through $\im(\xi(\hkc)) =
0$ and $\xi$ is given by Eq.~\eqref{eq:42}.  This behavior is
analogous to the supercritical Hopf bifurcation for the macrospin,
studied in the previous section.

In this section, we show that weakly nonlinear modulations of the
Slonczewski mode are no longer stable when $\hk > \hkc > 1$. This
behavior is analogous to the switching exhibited by the macrospin for
$\hk > \hkcm$ and $\sigma$ above threshold. While we are unable to
analytically follow the dynamics of the instability, numerical
simulations such as the droplet birth sequence shown in
Fig.~\ref{fig:droplet_birth} demonstrate that the magnetization
reverses inside the nanocontact and develops into a dissipative
droplet.

For the stability analysis, we seek a multiple scales solution
representing a modulation of the Slonczewski mode in the form
\begin{equation}
  \label{eq:33}
  \begin{split}
    u(\rv,t) = e^{i(h_0 + \hk - 1)t} e^{i \ws t} [ &\eps A(T) f(\rho)
    \\
    &+ \eps^3 u_1(\rv,T) + \cdots ], \\
    \sigma = \sis + \eps^2 \sigma_1 \qquad \quad \,&
  \end{split}
\end{equation}
where $0 < \eps \ll 1$ is the amplitude of the mode at the origin,
which is modulated by $A(T)$ with $T = \eps^2 t$ the ``slow'' time,
$\ws$ is the frequency of the Slonczewski mode $f(\rho)$ with
threshold current $\sis$, and $\sigma_1$ represents a deviation from
$\sis$.  The explicit form for $f$ and the implicit equations for
$\ws$ and $\sis$ are given in Appendix \ref{sec:modul-inst}.

\begin{figure}
  \centering
  \includegraphics[width=0.8\columnwidth]{
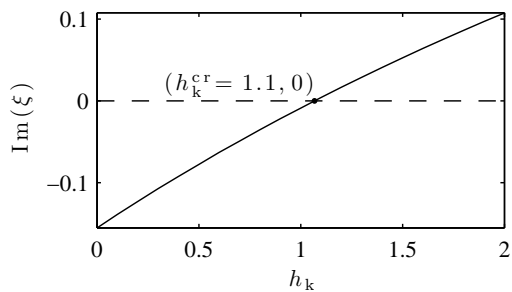}
  \caption{Modulation parameter $\im(\xi)$ as a function of $\hk$.
    When $\im(\xi) > 0$, the weakly nonlinear Slonczewski mode is
    modulationally unstable.  Parameter values are $\alpha = 0.03$,
    $\rho_* = 5.24$, $h_0 = 1.8$, $\nu = 0.26$.}
  \label{fig:xi_hk}
\end{figure}

\begin{figure}
  \centering
  \includegraphics[width=0.8\columnwidth]{
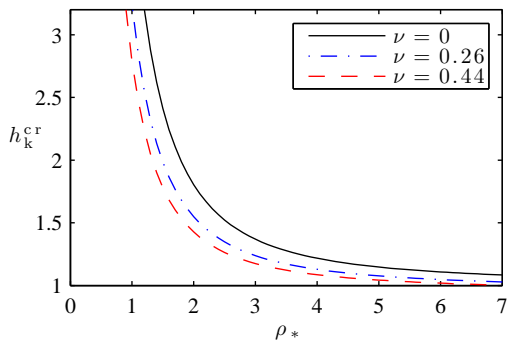}
  \caption{The critical anisotropy field $\hkc$ as a function of
    contact radius $\rho_*$ for various spin torque asymmetries $\nu$.
    Other parameters are $\alpha = 0.03$ and $h_0 = 1.8$.}
  \label{fig:hkc}
\end{figure}

Invoking a solvability condition at $\mathcal{O}(\eps^3)$,
Eq.~\eqref{eq:39} gives the nonlinear amplitude equation
\begin{equation}
  \label{eq:40}
  i \frac{d A}{d T} = i \frac{\sigma_1}{1 + \nu} \zeta A + \xi |A|^2 A,
\end{equation}
with complex linear and nonlinear coefficients $\zeta$ and $\xi$.
There is a time-periodic solution of Eq.~(\ref{eq:40}) for a specific
choice of $\sigma_1 \in \R$
\begin{equation}
  \label{eq:43}
  \begin{split}
    A(T) &= e^{i\wsp T}, \\
    \sigma_1 = -(1 + \nu) \frac{\im (\xi)}{\re (\zeta)}, &\quad \wsp = -
    \re (\xi) - \frac{\im (\xi) \im (\zeta)}{\re (\zeta)} .
  \end{split}
\end{equation}
This solution represents the nonlinear frequency shift $\eps^2 \wsp$
to the Slonczewski frequency $\ws$.  The stability analysis in
appendix \ref{sec:modul-inst} shows that the solution (\ref{eq:43}) is
unstable when $\im(\xi) > 0$.  An explicit formula for $\xi$ is given
in Eq.~(\ref{eq:42}).  We evaluate the integrals numerically and plot
$\im(\xi)$ as a function of $\hk$ for specific parameter values in
Fig.~\ref{fig:xi_hk}.
There is a critical value of the anisotropy field $\hkc$ satisfying
\begin{equation}
  \label{eq:59}
  \im[\xi(\hkc)] = 0,
\end{equation}
above which the weakly nonlinear Slonczewski mode is modulationally
unstable.  In other words, weak modulations of the Slonczewski mode
will grow exponentially in time when $\hk > \hkc$.  Figure
\ref{fig:hkc} shows the dependence of $\hkc = \hkc(\rho_*,\nu,h_0)$,
computed numerically by solving Eq.~\eqref{eq:59} with
Eq.~\eqref{eq:42}, as the contact radius and spin torque asymmetry are
varied  and $h_0 = 1.8$.  We see that $\hkc$ is
strictly greater than one and that larger perpendicular anisotropy is
required to enable the nucleation of a droplet in smaller nanocontact
devices. For a given contact size, larger spin torque asymmetry
permits nucleation of a droplet at smaller values of $\hk$.

We have also investigated the dependence of $\hkc$ on the applied
field magnitude $h_0$.  Since $h_0$ appears  in $\xi$
only multiplied by $\alpha$, we have $\pd{}{h_0}
\hkc(\rho_*,\nu,h_0) = \mathcal{O}(\alpha)$, which is small.  This is
confirmed by numerical calculations of the $\hkc$ dependence on $h_0
\ge 0$ for the values of $\rho_*$ and $\nu$ plotted in
Fig.~\ref{fig:hkc}.  We find that $\hkc$ varies by at most 3 \% for
$h_0 \in [0,1.8]$ with $\pd{}{h_0} \hkc(\rho_*,\nu,h_0) < 0$.
Importantly, the lower bound for $\hkc$ is preserved:
$\hkc(\rho_*,\nu,h_0) > 1$ when $h_0 \ge 0$.  This behavior stands in
stark contrast to the macrospin result for $\hkcm < 1$ in
Eq.~\eqref{eq:4} that strongly depends on the applied field.

From Eq.~(\ref{eq:43}), we see that $\sigma_1$ is negative when
$\im(\xi)/\re(\zeta) > 0$.  Numerically, we find that $\re(\zeta) >
0$, so that $\sigma_1 < 0$ when $\hk > \hkc$, which corresponds
precisely to the instability criterion.  Furthermore, when $\sigma_1 <
0$, any finite-amplitude excitation will tend to grow, thus the
time-periodic solution is  unstable.
This behavior is similar to the macrospin model discussed in the
previous section where, for $\hk > \hkc$, a small amplitude periodic
solution did not exist.

Numerical simulations of Eq.~(\ref{eq:1}) confirm the foregoing
analysis, even in the non-symmetric cases with $\theta_0 > 0$, $\thf >
0$, and $\hoe \ne 0$.  There is a critical value of the current above
which large amplitude dynamics ensue.  When $\hk$ is above $\hkc$, we
find that the magnetization inside the nanocontact reverses to form a
localized, coherently precessing, fully nonlinear magnetic solitary
wave.  We identify this solitary wave with the dissipative droplet
soliton found in the asymptotic analysis of Sec. \ref{sec:diss-dropl-solit}.

Due to the numerically robust formation of the droplet for a variety
of initial data and across a large parameter regime, we view it as a
global attractor.  As long as the current is above threshold, a
droplet is observed to nucleate.

\subsection{Hysteresis}
\label{sec:hysteresis}

We have shown in this section that a dissipative droplet will form
when the current exceeds the threshold for instability of small
amplitude waves inside the contact.  In the high symmetry case of
Eq.~(\ref{eq:53}) and $\thf = 0$, the threshold is the Slonczewski
critical current $\sis$, which is plotted in
Fig.~\ref{fig:droplet_f_I_oe} (triangle) along with the numerically
calculated threshold current from micromagnetic simulations that
incorporate the Oersted field and a canted fixed layer ($\sigma =
0.156$, vertical line).  We observe that the minimum droplet
sustaining current is below the threshold for droplet nucleation.
This suggests a hysteretic effect, whereby a droplet can be nucleated
at a current above threshold and remains stable when followed by an
adiabatic decrease of the current below threshold.  We have performed
this experiment numerically for the perturbed droplets of
Fig.~\ref{fig:droplet_f_I_oe}.
We use an already nucleated droplet (at $\sigma = 0.211$) 
as the initial condition for a new simulation with $\sigma = 0.151$,
below threshold.  We find that sufficiently close to, but below,
threshold, the droplet's frequency slightly increases but remains
stable.  For further decrease of the current to $\sigma = 0.136$,
however, the droplet undergoes a drift instability similar to behavior
shown in Fig.~\ref{fig:droplet_drift}.  Since the current is below
threshold a new droplet does not form.

\section{Discussion}
\label{sec:discussion}

The Slonczewski form for the spin torque term we are considering here,
Eq.~(\ref{eq:1}), was derived strictly for the case of spatially
uniform magnetization distributions
\cite{slonczewski_current-driven_1996}.  Nonlocal generalizations of
the spin torque term to nonuniform magnetization distributions have
been derived in the small amplitude regime
\cite{stiles_phenomenological_2004} and applied to single layer
nanocontact simulations \cite{hoefer_model_2008}.  Because the spin
torque and damping terms are treated as perturbations giving rise to a
dissipative droplet solution, we expect that the specific form for the
spin torque term, and the damping term for that matter, will not yield
qualitative changes in the structure of the droplet.  The asymptotic
analysis presented here is applicable to an arbitrary spin torque term
as long as its net effect is to oppose the inherent damping in the
system.

Due to the symmetries in the problem when $\theta_0 = \thf \approx 0$,
one might expect the Oersted field to prefer the excitation of a
localized vortex structure; e.g., a topological soliton
\cite{kosevich_magnetic_1990}.  We observe no such excitations in our
simulations.  Two-dimensional, conservative topological solitons
  are stable but they have higher energy than the conservative
  non-topological soliton \cite{kosevich_magnetic_1990}, providing one
  possible explanation for this behavior.  Another possible
  explanation lies in the form of the spin torque term.  In contrast
to the \emph{local} form for the spin torque considered here,
single-layer nanocontact simulations that incorporate a
\emph{nonlocal} spin torque have demonstrated the formation and
numerical stability of a precessional vortex in high-symmetry
configurations without perpendicular anisotropy
\cite{hoefer_model_2008}.  A more realistic model of the spin torque
may favor a topological soliton in certain cases.

Unfortunately, we as yet have no detailed physical understanding of
either the dissipative droplet asymmetry or the drift instability in
the presence of the combined effects of an Oersted field and a tilted
applied field.  In part, our lack of understanding stems from the
limited theory for the spatial propagation of the droplet
\cite{piette_localized_1998,cooper_solitary_1998}.  Extension of the
theory presented here to cases where the droplet is accelerated by
field gradients and other such forces may provide more physical
insight to aid in understanding why the droplet is subjected to
displacing forces.

The physical appearance of the droplet is reminiscent of the magnetic
bubble that was once the subject of intense investigation as a
possible alternative to ferrite-core computer memory
\cite{nielsen_bubble_1976}. Indeed, we can identify the zero-frequency
droplet as a topologically trivial magnetic bubble
\cite{komineas_topology_1996} with a winding number of zero, though in
the theory presented here, the droplet frequency approaches zero only
in the limit of infinitely large diameter. However, we expect that the
droplet will converge to the bubble structure at a finite diameter,
contingent on the inclusion of non-local magnetostatic energy in the
calculation; even when considering free layer films as thin as 3 nm,
such a term will eventually overcome the positive contribution of
exchange energy to the oscillator frequency, stabilizing the droplet
as a static structure. In this sense, then, we can think of the
conservative droplet soliton, a delicate balance between exchange and
perpendicular anisotropy, as a dynamically collapsing bubble, and the
dissipative droplet soliton as an imminently collapsing bubble that is
critically stabilized by the localized injection of spin torque.

\section{Conclusion}
\label{sec:conclusion}

We have derived equations for a dissipative droplet soliton through an
asymptotic analysis of the Landau-Lifshitz-Slonczewski equation for a
point-contact spin torque oscillator with perpendicular anisotropy in
the free layer. The droplet soliton is a localized, dynamic, solitary
wave solution consisting of partially reversed magnetization directly
under the contact and a zone of large amplitude precession in a region
bounding the reversed magnetization.  The diameter of the precessing
boundary is approximately the contact diameter.  The droplet frequency
is always strictly less than the ferromagnetic resonance frequency for
the film, and is also a monotonically decreasing function of droplet
diameter. The balance between spin torque and damping required to
sustain the droplet determines the relation between bias current and
frequency. The instability of small amplitude solutions that leads to
formation of the reversed domain in the droplet requires a minimum
perpendicular anisotropy that is a function of the contact radius and
spin torque asymmetry.

Some of the unique, identifying properties of the dissipative droplet
that could be observed experimentally include
\begin{itemize}
\item The droplet's frequency is well below the ferromagnetic
  resonance frequency.
\item Sufficiently far from the minimum sustaining current, the
  droplet's frequency has a weak dependence on current.
\item The droplet may manifest a drift instability, which would reveal
  itself as a transitory cessation in ac oscillations until the
  droplet again forms.
\item Hysteresis in current is expected, unless a drift instability
  results in a finite droplet lifetime.
\end{itemize}

We have investigated the nucleation and stability of the dissipative
droplet soliton through numerical simulations. We find that droplet
formation begins once the current in the point-contact is sufficient
to instigate the small amplitude Slonczewski mode, characterized by
spin waves that radiate away from the point-contact. For sufficiently
strong perpendicular anisotropy, this mode is subject to a
modulational instability and rapidly evolves into the reversed
magnetization profile of the droplet soliton. We find that the droplet
is stable in certain parameter regimes with regard to the
inhomogeneous Oersted field and to variations in spin torque asymmetry
and applied field angle.  Finally, the droplet is subject to a drift
instability that is a complicated function of the parameters employed
in this theory.

\appendix

\section{Modulational Instability of Slonczewski Mode}
\label{sec:modul-inst}

Here we provide the details of our stability analysis for small
amplitude, modulated waves excited in a nanocontact.

Inserting the ansatz~(\ref{eq:33}) into Eq.~(\ref{eq:32}) and
considering the leading order behavior in $\eps$ gives the linear
Slonczewski eigenmode $f(\rho)$ satisfying
\cite{slonczewski_excitation_1999}
\begin{equation}
  \label{eq:34}
  \begin{split}
    L_0 f \equiv &~(1 + i \alpha) (f'' + \frac{1}{\rho} f') - i \alpha
    (h_0 + \hk - 1) f \\
    &+ i \frac{\sis V(\rho)}{1 + \nu} f + \ws f = 0 .
  \end{split}
\end{equation}
Slonczewski considered both $\ws$ and $\sis$ as eigenvalues for this
equation and found the $C^1[0,\infty)$ solution
\begin{equation*}
  f(\rho) = \left \{
    \begin{array}{cc}
      J_0(\ki \rho), & 0 \le \rho \le \rho_* \\
      c H_0^{(1)}(\ko \rho), & \rho_* < \rho
    \end{array}
  \right ., \quad c = \frac{J_0(\ki \rho_*)}{H_0^{(1)}(\ko \rho_*)} ,
\end{equation*}
where $J_0$ is a Bessel function and $H_0^{(1)}$ is a Hankel function.
The inner and outer wavenumbers $\ki$, $\ko$ are
\begin{equation*}
  \begin{split}
    \ki &= \sqrt{\frac{\ws - i[\alpha(h_0 + \hk - 1) -
        \sis/(1+\nu)]}{1 +
        i \alpha}}, \\
    \ko &= -\sqrt{\frac{\ws - i\alpha(h_0 + \hk - 1)}{1 + i \alpha}} .
  \end{split}
\end{equation*}
Since $|f(\rho)| \sim C e^{-\im (\ko) \rho}/\sqrt{\rho}$ for $\rho \gg
\rho_*$, the sign of $\ko$ has been chosen so that $\im(\ko) > 0$, and
$f$ experiences exponential decay due to material damping $\alpha >
0$.  The decay length is weak compared to the contact radius.  The two
real eigenvalues $\ws$ and $\sis$ are determined by solving the
complex valued transcendental equation
\begin{equation*}
  \ki H_0^{(1)}(\ko \rho_*) J_1(\ki \rho_*) = \ko H_1^{(1)}(\ko
  \rho_*) J_0(\ki \rho_*) ,
\end{equation*}
which results from continuity of the first derivative of $f$.

Continuing the asymptotic analysis to the next order gives the
following nonhomogeneous equation
\begin{equation}
  \label{eq:38}
  \begin{split}
    L_0 u_1& = R_1 \equiv -i \frac{d A}{d T} f + i\frac{\sigma_1
      V(\rho)}{1 + \nu} A f\\
    + \frac{1}{2} |A|^2 A \Bigg \{ & \Big [\hk - 1 + i\alpha(h_0 + \hk
    - 1)
    - i \frac{\sis V(\rho)}{(1 + \nu)^2} \Big ]|f |^2 f \\
    &+ f \Big [\frac{d^2}{d\rho^2} + \frac{1}{\rho} \frac{d}{d\rho} \Big ] |f|^2 \\
    &- |f|^2 \Big [f'' + \frac{1}{\rho} f' \Big ] + 2 i \alpha | f'
    |^2 f \Bigg \} .
  \end{split}
\end{equation}
Since the kernel of the adjoint of $L_0$ is spanned by $f^*$ ($^*$
denotes complex conjugation), we invoke the solvability condition of
Eq.~(\ref{eq:38})
\begin{equation}
  \label{eq:39}
  \int_0^\infty f(\rho) R_1(\rho) \rho d \rho = 0 ,
\end{equation}
to determine the dynamical equation for $A(T)$ in Eq.~(\ref{eq:40}).
The complex valued linear and nonlinear coefficients $\zeta$ and $\xi$
are
\begin{align*}
  \zeta =&~ \frac{1 + [J_0(\ki \rho_*)/J_1(\ki \rho_*)]^2}{1 -
    [\ki/\ko]^2},
\end{align*}
and
\begin{widetext}
  \begin{equation}
    \label{eq:42}
    \xi = \frac{{\displaystyle [\hk - 1 + i \alpha(h_0 + \hk - 1)] \int_0^\infty
      |f|^2 f^2 \rho d \rho - i \frac{\sis}{(1 + \nu)^2}
      \int_0^{\rho_*} |f|^2 f^2 \rho d\rho  - (1 - 2i \alpha)
      \int_0^\infty |f'|^2 f^2 \rho d \rho}}{\rho_*^2 J_1(\ki
      \rho_*)^2[1 - (\ki/\ko)^2]} .
  \end{equation}
\end{widetext}

We now perform a linear stability analysis of the time-periodic
solution $A(T) = e^{i \wsp T}$ in Eq.~(\ref{eq:43}) by inserting
\begin{equation*}
  A(T) = e^{i \wsp T}( 1 + v + i w), \quad v, w \in \R, \quad |v|, |w|
  \ll 1,
\end{equation*}
into Eq.~(\ref{eq:40}).  Keeping only the terms linear in $v$ and $w$
gives the simple, decoupled dynamical system
\begin{equation*}
  \frac{dv}{dT} = 2 \im (\xi)v, \quad \frac{dw}{dT} = -2 \re(\xi) v ,
\end{equation*}
which experiences exponential growth when $\im(\xi) > 0$.

\section{Micromagnetic Computational Modeling}
\label{sec:numerical-method}

The numerical method we have used to solve Eq.\ (\ref{eq:1}) is
similar to the method tersely presented in
Ref.~\onlinecite{hoefer_model_2008} but adapted to a trilayer
nanocontact.  In this Appendix, we describe our method in detail.  In
brief, we use a polar coordinate grid and cylindrical magnetization
basis.  Angular derivatives are computed by use of a pseudospectral,
Fourier method while radial derivatives on a nonuniform grid are
computed by use of finite differences.  For time-stepping, we use an
explicit, hybrid $2^\textrm{nd}/3^\textrm{rd}$-order Runge-Kutta time
stepper with error control, \texttt{ode23} from
Matlab$^\textrm{\textregistered}$.

The polar coordinate system is a particularly efficient and accurate
choice for nanocontact simulations.  The discretization we use is
non-uniform in radius (``inner'' and ``outer'' grids)
\begin{equation}
  \label{eq:64}
  \begin{split}
    \rho_n = q(n) \equiv \rho_* \int_0^{n} \Big \{ &(\dro - \dri)
    \left[ \tanh \left (\frac{\xi
          - \hat{n}}{w} \right) + 1 \right] \\
    &+ \dri \Big \} \,d\xi ,
  \end{split}
\end{equation}
where $n = 1,\ldots,N$, and $\hat{n}$, $w$ are parameters determining
the location and width of the smooth change from the fine inner grid
spacing $\dri$ to the coarser outer grid spacing $\dro > \dri$.  We
typically have $\rho_1 \approx \dri$.  For the angular, pseudospectral
discretization, we expand the cylindrical magnetization basis in a
truncated Fourier series with $M$ Fourier modes
\begin{equation}
  \label{eq:60}
  \begin{split}
    &\m(\rho_n,\varphi,t) \approx \\
    &\sum_{k=-M/2}^{M/2-1} e^{i k \varphi} \left [
      \widehat{m}_{n,k}^{(\rho)}(t) \rhov +
      \widehat{m}_{n,k}^{(\varphi)}(t) \vf +
      \widehat{m}_{n,k}^{(z)}(t) \z \right ] .
  \end{split}
\end{equation}
The transformation from an equispaced angular grid $\varphi_k = -\pi +
(k-1) 2\pi/M$ with magnetization coefficients
$m_{n,k}^{(\rho,\varphi,z)}(t)$ evaluated at the discrete grid
$(\rho_n,\varphi_k)$, to the Fourier representation in
Eq.~(\ref{eq:60}) is achieved by use of the fast Fourier transform.

The advantage of this discretization is that we can solve on a uniform
computational grid while the physical grid is clustered in and around
the nanocontact, where the majority of the fine scale dynamics occur.
The outer grid supports the propagation of spin waves of the
appropriate wavelength away from the nanocontact.  In order to
simulate an infinite domain, we choose a finite domain large enough,
$L \equiv \rho_N \approx 30 \rho_*$, so that spurious waves are
naturally damped to a negligible amplitude.  Then their reflection off
the boundary does not affect the strongly localized dynamics near the
nanocontact.

Computing the vector Laplacian $\nabla^2 \m$ is the computationally
intensive portion of the algorithm.  Angular derivatives are
approximated by multiplication in Fourier space $\pd{}{\varphi} \to
ik$.  The approximation of radial derivatives using finite differences
requires some care, especially near the origin.  We discuss the
details now.  

Due to the non-uniform radial grid (\ref{eq:64}), radial derivatives
in computational space require appropriate factors of $q$ and its
derivatives.  For example
\begin{equation*}
  \pd{}{\rho} \to \frac{1}{q'} \pd{}{n}, \quad \pdd{}{\rho} \to
  \frac{1}{q'^3} \left ( q'
    \pdd{}{n} - q'' \pd{}{n} \right ) .
\end{equation*}
The radial derivatives in computational space are approximated using
$6^\textrm{th}$ order finite differences.  The boundary condition at
the artificial boundary $\rho = L$ for Eq.\ \eqref{eq:1} is the
Neumann condition $\pd{\widehat{m}^{(\rho,\varphi,z)}_N}{\rho}(t) =
0$, which is implemented by use of a standard ghost point method and
one-sided differences.  For radial derivatives near the origin at $n =
1,2$, we define symmetry conditions to obtain
$\widehat{m}^{(\rho,\varphi,z)}_{-n,k}(t)$ by taking $\varphi \to
\varphi + \pi$:
\begin{equation}
  \label{eq:61}
  \begin{split}
    \widehat{m}^{(\rho)}_{n,k}(t) e^{i k\varphi} \rhov &= (-1)^{k+1}
    \widehat{m}^{(\rho)}_{n,k}(t)
    e^{ik(\varphi + \pi)} (-\rhov ) \\
    &\equiv (-1)^{k+1} \widehat{m}^{(\rho)}_{-n,k}(t) e^{i k\varphi} \rhov ,\\
    \widehat{m}^{(\varphi)}_{n,k}(t) e^{i k\varphi} \vf &= (-1)^{k+1}
    \widehat{m}^{(\varphi)}_{n,k}(t)
    e^{ik(\varphi + \pi)} (-\vf ) \\
    &\equiv (-1)^{k+1} \widehat{m}^{(\varphi)}_{-n,k}(t) e^{i k\varphi} \vf ,\\
    \widehat{m}^{(z)}_{n,k}(t) e^{i k\varphi} \z &= (-1)^{k}
    \widehat{m}^{(z)}_{n,k}(t)
    e^{ik(\varphi + \pi)} \z \\
    &\equiv (-1)^{k} \widehat{m}^{(z)}_{-n,k}(t) e^{i k\varphi} \z ,\\
  \end{split}
\end{equation}
where we have used $\rhov \to -\rhov$ and $\vf \to -\vf$ when $\varphi
\to \varphi + \pi$.  Therefore,
$\widehat{m}^{(\rho,\varphi)}_{n,k}(t)$ are even/odd functions of $n$
as $k$ is odd/even while $\widehat{m}^{(z)}_{n,k}(t)$ is even/odd as
$k$ is even/odd.  At $\rho = 0$, we take (see
Ref.~\onlinecite{iserles_2008})
\begin{equation}
  \label{eq:63}
  \begin{split}
    \widehat{m}^{(\rho,\varphi)}_{0,k}(t) &= 0, \quad |k| \ne 1, \\
    \widehat{m}^{(z)}_{0,k}(t) &= 0, \quad k \ne 0, \\
    \pd{\widehat{m}^{(\rho,\varphi)}_{0,\pm 1}}{\rho}(t) &= 0, \quad
    \pd{\widehat{m}^{(z)}_{0,0}(t)}{\rho} = 0 .
  \end{split}
\end{equation}
The derivative conditions in Eq.~(\ref{eq:63}) are approximated by use
of one-sided differences to obtain an estimate of $\widehat{m}_{0,\pm
  1}^{(\rho,\varphi)}(t)$ and $\widehat{m}_{0,0}^{(z)}(t)$.  The
symmetries in Eq.~(\ref{eq:61}) and the conditions in
Eq.~(\ref{eq:63}) enable the use of centered finite differences, even
near the origin.  An explicit, Runge-Kutta
$2^\textrm{nd}/3^\textrm{rd}$-order time-stepping method is used to
advance the discretized version of equation \eqref{eq:1} forward in
time while renormalizing the magnitude of $\m$ after every time-step
to preserve the constraint $|\m| = 1$.  To avoid severe time-step
restrictions due to the small grid spacing near the origin ($\approx
2\pi \rho_n/M$), we apply a smooth, radial grid dependent angular mask
that effectively reduces the number of angular modes at $\rho_n$ from
$M$ to $M_n = 2\pi/k_n$:
\begin{equation*}
  g_n(k) = \frac{1}{2} + \frac{1}{2}\tanh\left(\frac{k_n - |k|}{\Delta
      k}  \right) .
\end{equation*}
The mask's parameters are the radial grid dependent wavenumber cutoffs
$k_n$, $n=1,\ldots, N$ and the cutoff width $\Delta k$.  Application
of the mask at every time step filters out numerically induced small
wavelengths near the origin \cite{fornberg_pseudospectral_1995}.  The
cutoffs $k_n$ are chosen so that the approximate grid spacing $2\pi
\rho_n/M_n \sim \dri$,  hence the grid near the
  origin has an effective spacing of $\dri$.  The mask applied to
$\widehat{m}^{(z)}_{n,k}(t)$ takes the form
\begin{equation*}
  \mathcal{G}^{(z)} \left \{ \widehat{m}^{(z)}(t) \right \}_{n,k} =
  g_n(k) \widehat{m}^{(z)}_{n,k}(t) .
\end{equation*}
Care must be taken when applying the mask to the in-plane Fourier
coefficients $\widehat{m}^{(\rho,\varphi)}_{n,k}(t)$ because $\rhov$
and $\vf$ depend on the grid location.  We use
\begin{equation*}
  \begin{split}
    \mathcal{G}^{(\rho)} \left \{ \widehat{m}^{(\rho)}(t) \right
    \}_{n,k}& = 
  \frac{1}{2} \Big \{ [g_{n}(k-1) + g_{n}(k+1) ]
    \widehat{m}^{(\rho)}_{n,k}(t) \\
    &+ i [ -g_{n}(k-1) + g_{n}(k+1) ]
    \widehat{m}^{(\varphi)}_{n,k}(t)  \Big \}, \\
    \mathcal{G}^{(\varphi)} \left \{ \widehat{m}^{(\varphi)}(t) \right
    \}_{n,k}& = 
  \frac{1}{2} \Big \{ i [g_{n}(k-1) - g_{n}(k+1) ]
    \widehat{m}^{(\rho)}_{n,k}(t) \\
    &+ [g_{n}(k-1) + g_{n}(k+1) ]
    \widehat{m}^{(\varphi)}_{n,k}(t)  \Big \} .
  \end{split}
\end{equation*}

Numerical parameters we use are: $\dri= 0.048 \rho_*$, $\dro = 0.25
\rho_*$, $w = 10$, $\hat{n} = 126$, $M = 32 - 128$, $N = 222$, $L = 30
\rho_*$, $k_1 = 4$, and $\Delta k = 1$.  We find no significant change
in the presented results for more accurate grids and filtering
parameters.

In order to nucleate a droplet, we use initial conditions that are
saturated in the $\z$ direction with a current that is above
threshold.  The small amplitude instability investigated in
Sec.~\ref{sec:excitation-droplet} leads to the formation of a droplet.
For the computation of the frequencies of the droplet in
Fig.~\ref{fig:droplet_f_I_oe} found from micromagnetics, we take the
Fourier transform of the spatially averaged magnetization time series
from the simulation and extract the frequency with the largest power.
The frequency resolution is finer than the size of the dots in
Fig.~\ref{fig:droplet_f_I_oe}.


\bibliographystyle{apsrev}

\begin{thebibliography}{32}
\expandafter\ifx\csname natexlab\endcsname\relax\def\natexlab#1{#1}\fi
\expandafter\ifx\csname bibnamefont\endcsname\relax
  \def\bibnamefont#1{#1}\fi
\expandafter\ifx\csname bibfnamefont\endcsname\relax
  \def\bibfnamefont#1{#1}\fi
\expandafter\ifx\csname citenamefont\endcsname\relax
  \def\citenamefont#1{#1}\fi
\expandafter\ifx\csname url\endcsname\relax
  \def\url#1{\texttt{#1}}\fi
\expandafter\ifx\csname urlprefix\endcsname\relax\def\urlprefix{URL }\fi
\providecommand{\bibinfo}[2]{#2}
\providecommand{\eprint}[2][]{\url{#2}}

\bibitem[{\citenamefont{Akhmediev and
      Ankiewicz}(2008{\natexlab{a}})}]{akhmediev_dissipative_2008}
  \bibinfo{author}{\bibfnamefont{N.~N.} \bibnamefont{Akhmediev}}
  \bibnamefont{and}
  \bibinfo{author}{\bibfnamefont{A.}~\bibnamefont{Ankiewicz}},
  \emph{\bibinfo{title}{Dissipative solitons: from optics to biology
      and medicine}}, Lecture notes in physics,
  \textbf{\bibinfo{volume}{751}} (\bibinfo{publisher}{Springer},
  \bibinfo{address}{Berlin}, \bibinfo{year}{2008}).

\bibitem[{\citenamefont{Ivanov and Kosevich}(1977)}]{ivanov_bound-states_1977}
\bibinfo{author}{\bibfnamefont{B.~A.} \bibnamefont{Ivanov}} \bibnamefont{and}
  \bibinfo{author}{\bibfnamefont{A.~M.} \bibnamefont{Kosevich}},
  \bibinfo{journal}{Zhurnal Eksperimentalnoi I Teoreticheskoi Fiziki}
  \textbf{\bibinfo{volume}{72}}, \bibinfo{pages}{2000} (\bibinfo{year}{1977}).

\bibitem[{\citenamefont{Kosevich et~al.}(1990)\citenamefont{Kosevich, Ivanov,
  and Kovalev}}]{kosevich_magnetic_1990}
\bibinfo{author}{\bibfnamefont{A.~M.} \bibnamefont{Kosevich}},
  \bibinfo{author}{\bibfnamefont{B.~A.} \bibnamefont{Ivanov}},
  \bibnamefont{and} \bibinfo{author}{\bibfnamefont{A.~S.}
  \bibnamefont{Kovalev}}, \bibinfo{journal}{Phys. Rep.}
  \textbf{\bibinfo{volume}{194}}, \bibinfo{pages}{117} (\bibinfo{year}{1990}).

\bibitem[{\citenamefont{Ablowitz and
      Segur}(1981)}]{ablowitz_solitons_1981}
  \bibinfo{author}{\bibfnamefont{M.~J.} \bibnamefont{Ablowitz}}
  \bibnamefont{and}
  \bibinfo{author}{\bibfnamefont{H.}~\bibnamefont{Segur}},
  \emph{\bibinfo{title}{Solitons and the inverse scattering
      transform}}, (\bibinfo{publisher}{{SIAM}},
  \bibinfo{address}{Philadelphia}, \bibinfo{year}{1981}).

\bibitem[{\citenamefont{Zvezdin and Popkov}(1983)}]{zvezdin_contribution_1983}
\bibinfo{author}{\bibfnamefont{A.~K.} \bibnamefont{Zvezdin}} \bibnamefont{and}
  \bibinfo{author}{\bibfnamefont{A.~F.} \bibnamefont{Popkov}},
  \bibinfo{journal}{Sov. Phys. {JETP}} \textbf{\bibinfo{volume}{57}},
  \bibinfo{pages}{350–355} (\bibinfo{year}{1983}).
\bibinfo{author}{\bibfnamefont{B.~A.} \bibnamefont{Kalinikos}},
  \bibinfo{author}{\bibfnamefont{N.~G.} \bibnamefont{Kovshikov}},
  \bibnamefont{and} \bibinfo{author}{\bibfnamefont{A.~N.}
  \bibnamefont{Slavin}}, \bibinfo{journal}{{JETP} Lett.}
  \textbf{\bibinfo{volume}{38}}, \bibinfo{pages}{413} (\bibinfo{year}{1983}).
\bibinfo{author}{\bibfnamefont{P.} \bibnamefont{DeGasperis}},
  \bibinfo{author}{\bibfnamefont{R.}~\bibnamefont{Marcelli}}, \bibnamefont{and}
  \bibinfo{author}{\bibfnamefont{G.}~\bibnamefont{Miccoli}},
  \bibinfo{journal}{Phys. Rev. Lett.} \textbf{\bibinfo{volume}{59}},
  \bibinfo{pages}{481} (\bibinfo{year}{1987}).
\bibinfo{author}{\bibfnamefont{B.~A.} \bibnamefont{Kalinikos}},
  \bibinfo{author}{\bibfnamefont{N.~G.} \bibnamefont{Kovshikov}},
  \bibnamefont{and} \bibinfo{author}{\bibfnamefont{A.~N.}
  \bibnamefont{Slavin}}, \bibinfo{journal}{Phys. Rev. B}
  \textbf{\bibinfo{volume}{42}}, \bibinfo{pages}{8658} (\bibinfo{year}{1990}).
\bibinfo{author}{\bibfnamefont{M.}~\bibnamefont{Chen}},
  \bibinfo{author}{\bibfnamefont{M.~A.} \bibnamefont{Tsankov}},
  \bibinfo{author}{\bibfnamefont{J.~M.} \bibnamefont{Nash}}, \bibnamefont{and}
  \bibinfo{author}{\bibfnamefont{C.~E.} \bibnamefont{Patton}},
  \bibinfo{journal}{Phys. Rev. Lett.} \textbf{\bibinfo{volume}{70}},
  \bibinfo{pages}{1707} (\bibinfo{year}{1993}).

\bibitem[{\citenamefont{Akhmediev and
  Ankiewicz}(2008{\natexlab{b}})}]{akhmediev__2008}
\bibinfo{author}{\bibfnamefont{N.}~\bibnamefont{Akhmediev}} \bibnamefont{and}
  \bibinfo{author}{\bibfnamefont{A.}~\bibnamefont{Ankiewicz}},
  \bibinfo{journal}{Lect. Notes Phys.} \textbf{\bibinfo{volume}{751}},
  \bibinfo{pages}{1} (\bibinfo{year}{2008}).

\bibitem[{\citenamefont{Slonczewski}(1996)}]{slonczewski_current-driven_1996}
\bibinfo{author}{\bibfnamefont{J.~C.} \bibnamefont{Slonczewski}},
  \bibinfo{journal}{J. Magn. Magn. Mater.}
  \textbf{\bibinfo{volume}{159}}, \bibinfo{pages}{L1} (\bibinfo{year}{1996}).

\bibitem[{\citenamefont{Berger}(1996)}]{berger__1996}
\bibinfo{author}{\bibfnamefont{L.}~\bibnamefont{Berger}},
  \bibinfo{journal}{Phys. Rev. B} \textbf{\bibinfo{volume}{54}},
  \bibinfo{pages}{9353} (\bibinfo{year}{1996}).

\bibitem[{\citenamefont{Slonczewski}(1999)}]{slonczewski_excitation_1999}
\bibinfo{author}{\bibfnamefont{J.~C.} \bibnamefont{Slonczewski}},
  \bibinfo{journal}{J. Magn. Magn. Mater.} \textbf{\bibinfo{volume}{195}},
  \bibinfo{pages}{L261} (\bibinfo{year}{1999}).

\bibitem[{\citenamefont{Myers et~al.}(1999)\citenamefont{Myers, Ralph, Katine,
  Louie, and Buhrman}}]{myers__1999}
\bibinfo{author}{\bibfnamefont{E.~B.} \bibnamefont{Myers}},
  \bibinfo{author}{\bibfnamefont{D.~C.} \bibnamefont{Ralph}},
  \bibinfo{author}{\bibfnamefont{J.~A.} \bibnamefont{Katine}},
  \bibinfo{author}{\bibfnamefont{R.~N.} \bibnamefont{Louie}}, \bibnamefont{and}
  \bibinfo{author}{\bibfnamefont{R.~A.} \bibnamefont{Buhrman}},
  \bibinfo{journal}{Science} \textbf{\bibinfo{volume}{285}},
  \bibinfo{pages}{867} (\bibinfo{year}{1999}).
\bibinfo{author}{\bibfnamefont{J.~A.} \bibnamefont{Katine}},
  \bibinfo{author}{\bibfnamefont{F.~J.} \bibnamefont{Albert}},
  \bibinfo{author}{\bibfnamefont{R.~A.} \bibnamefont{Buhrman}},
  \bibinfo{author}{\bibfnamefont{E.~B.} \bibnamefont{Myers}}, \bibnamefont{and}
  \bibinfo{author}{\bibfnamefont{D.~C.} \bibnamefont{Ralph}},
  \bibinfo{journal}{Phys. Rev. Lett.} \textbf{\bibinfo{volume}{84}},
  \bibinfo{pages}{3149} (\bibinfo{year}{2000}).

\bibitem[{\citenamefont{Kiselev et~al.}(2003)\citenamefont{Kiselev, Sankey,
  Krivotorov, Emley, Schoelkopf, Buhrman, and Ralph}}]{kiselev_microwave_2003}
\bibinfo{author}{\bibfnamefont{S.~I.} \bibnamefont{Kiselev}},
  \bibinfo{author}{\bibfnamefont{J.~C.} \bibnamefont{Sankey}},
  \bibinfo{author}{\bibfnamefont{I.~N.} \bibnamefont{Krivotorov}},
  \bibinfo{author}{\bibfnamefont{N.~C.} \bibnamefont{Emley}},
  \bibinfo{author}{\bibfnamefont{R.~J.} \bibnamefont{Schoelkopf}},
  \bibinfo{author}{\bibfnamefont{R.~A.} \bibnamefont{Buhrman}},
  \bibnamefont{and} \bibinfo{author}{\bibfnamefont{D.~C.} \bibnamefont{Ralph}},
  \bibinfo{journal}{Nature} \textbf{\bibinfo{volume}{425}},
  \bibinfo{pages}{380} (\bibinfo{year}{2003}).
\bibinfo{author}{\bibfnamefont{W.~H.} \bibnamefont{Rippard}},
  \bibinfo{author}{\bibfnamefont{M.~R.} \bibnamefont{Pufall}},
  \bibinfo{author}{\bibfnamefont{S.}~\bibnamefont{Kaka}},
  \bibinfo{author}{\bibfnamefont{S.~E.} \bibnamefont{Russek}},
  \bibnamefont{and} \bibinfo{author}{\bibfnamefont{T.~J.} \bibnamefont{Silva}},
  \bibinfo{journal}{Phys. Rev. Lett.} \textbf{\bibinfo{volume}{92}},
  \bibinfo{pages}{027201} (\bibinfo{year}{2004}).

\bibitem[{\citenamefont{Ralph and Stiles}(2008)}]{ralph_spin_2008}
\bibinfo{author}{\bibfnamefont{D.}~\bibnamefont{Ralph}} \bibnamefont{and}
  \bibinfo{author}{\bibfnamefont{M.}~\bibnamefont{Stiles}},
  \bibinfo{journal}{J. Magn. Magn. Mater.}
  \textbf{\bibinfo{volume}{320}}, \bibinfo{pages}{1190} (\bibinfo{year}{2008}).

\bibitem[{\citenamefont{Silva and Rippard}(2008)}]{silva_developments_2008}
\bibinfo{author}{\bibfnamefont{T.}~\bibnamefont{Silva}} \bibnamefont{and}
  \bibinfo{author}{\bibfnamefont{W.}~\bibnamefont{Rippard}},
  \bibinfo{journal}{J. Magn. Magn. Mater.}
  \textbf{\bibinfo{volume}{320}}, \bibinfo{pages}{1260} (\bibinfo{year}{2008}).
\bibinfo{author}{\bibfnamefont{J.~A.} \bibnamefont{Katine}} \bibnamefont{and}
  \bibinfo{author}{\bibfnamefont{E.~E.} \bibnamefont{Fullerton}},
  \bibinfo{journal}{J. Magn. Magn. Mater.} \textbf{\bibinfo{volume}{320}},
  \bibinfo{pages}{1217} (\bibinfo{year}{2008}).

\bibitem[{\citenamefont{Slavin and Tiberkevich}(2005)}]{slavin_spin_2005}
\bibinfo{author}{\bibfnamefont{A.}~\bibnamefont{Slavin}} \bibnamefont{and}
  \bibinfo{author}{\bibfnamefont{V.}~\bibnamefont{Tiberkevich}},
  \bibinfo{journal}{Phys. Rev. Lett.} \textbf{\bibinfo{volume}{95}},
  \bibinfo{pages}{237201} (\bibinfo{year}{2005}).

\bibitem[{\citenamefont{Schryer and Walker}(1974)\citenamefont{Schryer
      and Walker}}]{schryer_1974}
\bibinfo{author}{\bibfnamefont{N.~L.} \bibnamefont{Schryer}} \bibnamefont{and}
  \bibinfo{author}{\bibfnamefont{L.~R.} \bibnamefont{Walker}},
  \bibinfo{journal}{J. Appl. Phys.} \textbf{\bibinfo{volume}{45}},
  \bibinfo{pages}{5406} (\bibinfo{year}{1974}).
  \bibinfo{author}{\bibfnamefont{J.-Y.} \bibnamefont{Lee}},
  \bibinfo{author}{\bibfnamefont{K.-S.} \bibnamefont{Lee}},
  \bibinfo{author}{\bibfnamefont{S.} \bibnamefont{Choi}},
  \bibinfo{author}{\bibfnamefont{K.~Y.} \bibnamefont{Guslienko}},
  \bibnamefont{and} \bibinfo{author}{\bibfnamefont{S.-K.}
    \bibnamefont{Kim}}, \bibinfo{journal}{Phys. Rev. B}
  \textbf{\bibinfo{volume}{76}}, \bibinfo{pages}{184408}
  (\bibinfo{year}{2007}).

\bibitem[{\citenamefont{Nielsen}(1976)}]{nielsen_bubble_1976}
\bibinfo{author}{\bibfnamefont{J.}~\bibnamefont{Nielsen}},
  \bibinfo{journal}{{IEEE} Trans. Mag.} \textbf{\bibinfo{volume}{12}},
  \bibinfo{pages}{327} (\bibinfo{year}{1976}).
\bibinfo{author}{\bibfnamefont{E.~A.} \bibnamefont{Giess}},
  \bibinfo{journal}{Science} \textbf{\bibinfo{volume}{208}},
  \bibinfo{pages}{938} (\bibinfo{year}{1980}).

\bibitem[{\citenamefont{Guslienko et~al.}(2005)}]{guslienko_2005}
\bibinfo{author}{\bibfnamefont{K.~Y.} \bibnamefont{Guslienko}},
\bibinfo{author}{\bibfnamefont{W.} \bibnamefont{Scholz}},
\bibinfo{author}{\bibfnamefont{R.~W.} \bibnamefont{Chantrell}}, and
\bibinfo{author}{\bibfnamefont{V.} \bibnamefont{Novosad}},
  \bibinfo{journal}{Phys. Rev. B} \textbf{\bibinfo{volume}{71}},
  \bibinfo{pages}{144407} (\bibinfo{year}{2005}).
\bibinfo{author}{\bibfnamefont{V.}~\bibnamefont{Novosad}},
\bibinfo{author}{\bibfnamefont{F.~Y.}~\bibnamefont{Fradin}},
\bibinfo{author}{\bibfnamefont{P.~E.}~\bibnamefont{Roy}},
\bibinfo{author}{\bibfnamefont{K.~S.}~\bibnamefont{Buchanan}},
\bibinfo{author}{\bibfnamefont{K.~Y.}~\bibnamefont{Guslienko}}, \bibnamefont{and}
  \bibinfo{author}{\bibfnamefont{S.~D.}~\bibnamefont{Bader}},
  \bibinfo{journal}{Phys. Rev. B}
  \textbf{\bibinfo{volume}{72}}, \bibinfo{pages}{024455} (\bibinfo{year}{2005}).

\bibitem[{\citenamefont{Gioia and James}(1997)}]{gioia_micromagnetics_1997}
\bibinfo{author}{\bibfnamefont{G.}~\bibnamefont{Gioia}} \bibnamefont{and}
  \bibinfo{author}{\bibfnamefont{R.~D.} \bibnamefont{James}},
  \bibinfo{journal}{Proc. R. Soc. Lond. A} \textbf{\bibinfo{volume}{453}},
  \bibinfo{pages}{213} (\bibinfo{year}{1997}).

\bibitem[{\citenamefont{Hoefer et~al.}(2008)\citenamefont{Hoefer, Silva, and
  Stiles}}]{hoefer_model_2008}
\bibinfo{author}{\bibfnamefont{M.~A.} \bibnamefont{Hoefer}},
  \bibinfo{author}{\bibfnamefont{T.~J.} \bibnamefont{Silva}}, \bibnamefont{and}
  \bibinfo{author}{\bibfnamefont{M.~D.} \bibnamefont{Stiles}},
  \bibinfo{journal}{Phys. Rev. B} \textbf{\bibinfo{volume}{77}},
  \bibinfo{pages}{144401} (\bibinfo{year}{2008}).

\bibitem[{\citenamefont{Kevorkian and Cole}(1996)}]{kevorkian_multiple_1996}
\bibinfo{author}{\bibfnamefont{J.}~\bibnamefont{Kevorkian}} \bibnamefont{and}
  \bibinfo{author}{\bibfnamefont{J.~D.} \bibnamefont{Cole}},
  \emph{\bibinfo{title}{Multiple scale and singular perturbation methods}}
  (\bibinfo{publisher}{Springer}, \bibinfo{year}{1996}).

\bibitem[{\citenamefont{Berkov and
      Gorn}(2007)}]{berkov_magnetization_2007}
  \bibinfo{author}{\bibfnamefont{D.~V.} \bibnamefont{Berkov}}
  \bibnamefont{and} \bibinfo{author}{\bibfnamefont{N.~L.}
    \bibnamefont{Gorn}}, \bibinfo{journal}{Phys. Rev. B}
  \textbf{\bibinfo{volume}{76}}, \bibinfo{pages}{144414}
  (\bibinfo{year}{2007}).

\bibitem[{\citenamefont{Piette and Zakrzewski}(1998)}]{piette_localized_1998}
\bibinfo{author}{\bibfnamefont{B.}~\bibnamefont{Piette}} \bibnamefont{and}
  \bibinfo{author}{\bibfnamefont{W.~J.} \bibnamefont{Zakrzewski}},
  \bibinfo{journal}{Physica D}
  \textbf{\bibinfo{volume}{119}}, \bibinfo{pages}{314} (\bibinfo{year}{1998}).

\bibitem[{\citenamefont{Bazaliy et~al.}(2004)}]{bazaliy_2004}
  \bibinfo{author}{\bibfnamefont{Ya.~B.}~\bibnamefont{Bazaliy}},
  \bibinfo{author}{\bibfnamefont{B.~A.}~\bibnamefont{Jones}},
  \bibnamefont{and} \bibinfo{author}{\bibfnamefont{S.-C.}
    \bibnamefont{Zhang}}, \bibinfo{journal}{Phys.~Rev.~B}
  \textbf{\bibinfo{volume}{69}}, \bibinfo{pages}{0094421}
  (\bibinfo{year}{2004}).
  \bibinfo{author}{\bibfnamefont{S.}~\bibnamefont{Mangin}},
  \bibinfo{author}{\bibfnamefont{D.}~\bibnamefont{Ravelosona}},
  \bibinfo{author}{\bibfnamefont{J.~A.}~\bibnamefont{Katine}},
  \bibinfo{author}{\bibfnamefont{M.~J.}~\bibnamefont{Carey}},
  \bibinfo{author}{\bibfnamefont{B.~D.}~\bibnamefont{Terris}},
  \bibnamefont{and} \bibinfo{author}{\bibfnamefont{E.~E.}
    \bibnamefont{Fullerton}}, \bibinfo{journal}{Nat. Mater.}
  \textbf{\bibinfo{volume}{5}}, \bibinfo{pages}{210}
  (\bibinfo{year}{2006}).
  \bibinfo{author}{\bibfnamefont{C.}~\bibnamefont{Serpico}},
  \bibinfo{author}{\bibfnamefont{G.}~\bibnamefont{Bertotti}},
  \bibinfo{author}{\bibfnamefont{R.}~\bibnamefont{Bonin}},
  \bibinfo{author}{\bibfnamefont{M.}~\bibnamefont{d'Aquino}},
  \bibnamefont{and} \bibinfo{author}{\bibfnamefont{I.~D.}
    \bibnamefont{Mayergoyz}}, \bibinfo{journal}{J. Appl. Phys.}
  \textbf{\bibinfo{volume}{101}}, \bibinfo{pages}{09A507}
  (\bibinfo{year}{2007}).

\bibitem[{\citenamefont{Hoefer et~al.}(2005)\citenamefont{Hoefer, Ablowitz,
  Ilan, Pufall, and Silva}}]{hoefer_theory_2005}
\bibinfo{author}{\bibfnamefont{M.~A.} \bibnamefont{Hoefer}},
  \bibinfo{author}{\bibfnamefont{M.~J.} \bibnamefont{Ablowitz}},
  \bibinfo{author}{\bibfnamefont{B.}~\bibnamefont{Ilan}},
  \bibinfo{author}{\bibfnamefont{M.~R.} \bibnamefont{Pufall}},
  \bibnamefont{and} \bibinfo{author}{\bibfnamefont{T.~J.} \bibnamefont{Silva}},
  \bibinfo{journal}{Phys. Rev. Lett.} \textbf{\bibinfo{volume}{95}},
  \bibinfo{pages}{267206} (\bibinfo{year}{2005}).

\bibitem[{\citenamefont{Stiles et~al.}(2004)\citenamefont{Stiles, Xiao, and
  Zangwill}}]{stiles_phenomenological_2004}
\bibinfo{author}{\bibfnamefont{M.~D.} \bibnamefont{Stiles}},
  \bibinfo{author}{\bibfnamefont{J.}~\bibnamefont{Xiao}}, \bibnamefont{and}
  \bibinfo{author}{\bibfnamefont{A.}~\bibnamefont{Zangwill}},
  \bibinfo{journal}{Phys. Rev. B} \textbf{\bibinfo{volume}{69}},
  \bibinfo{pages}{054408} (\bibinfo{year}{2004}).

\bibitem[{\citenamefont{Cooper}(1998)}]{cooper_solitary_1998}
\bibinfo{author}{\bibfnamefont{N.~R.}~\bibnamefont{Cooper}},
  \bibinfo{journal}{Phys. Rev. Lett.}
  \textbf{\bibinfo{volume}{80}}, \bibinfo{pages}{4554} (\bibinfo{year}{1998})

\bibitem[{\citenamefont{Komineas and
  Papanicolaou}(1996)}]{komineas_topology_1996}
\bibinfo{author}{\bibfnamefont{S.}~\bibnamefont{Komineas}} \bibnamefont{and}
  \bibinfo{author}{\bibfnamefont{N.}~\bibnamefont{Papanicolaou}},
  \bibinfo{journal}{Physica D}
  \textbf{\bibinfo{volume}{99}}, \bibinfo{pages}{81} (\bibinfo{year}{1996}).

\bibitem[{\citenamefont{Iserles}(2008)}]{iserles_2008}
  \bibinfo{author}{\bibfnamefont{A.} \bibnamefont{Iserles}},
  \emph{\bibinfo{title}{A first course in the numerical analysis of
      differential equations}}, (\bibinfo{publisher}{Cambridge
    University Press},
  \bibinfo{address}{Cambridge}, \bibinfo{year}{2008}).

\bibitem[{\citenamefont{Fornberg}(1995)}]{fornberg_pseudospectral_1995}
  \bibinfo{author}{\bibfnamefont{B.}~\bibnamefont{Fornberg}},
  \bibinfo{journal}{{SIAM} J. Sci. Comp.}
  \textbf{\bibinfo{volume}{16}}, \bibinfo{pages}{1071}
  (\bibinfo{year}{1995}).

\end{thebibliography}

\end{document}